\documentclass[twocolumn]{aastex61}

\newcommand{\HII}{\ion{H}{2}}
\newcommand{\Msol}{$M_\odot$}
\newcommand{\kms}{\mbox{km~s$^{-1}$}}
\newcommand{\mjybm}{\mbox{mJy~beam$^{-1}$}}

\def\twco{\mbox{$^{12}$CO}}
\def\ttco{\mbox{$^{13}$CO}}

\shorttitle{A Quiescent Molecular Cloud in the LMC}
\shortauthors{Wong et al.}

\begin{document}

\title{ALMA Observations of a Quiescent Molecular Cloud in the Large Magellanic Cloud}

\correspondingauthor{Tony Wong}
\email{wongt@illinois.edu}

\author[0000-0002-7759-0585]{Tony Wong}
\affil{Department of Astronomy, University of Illinois, 
Urbana, IL 61801, USA}

\author{Annie Hughes}
\affil{Universit\'{e} de Toulouse, UPS-OMP, 31028 Toulouse, France}
\affil{CNRS, IRAP, Av. du Colonel Roche BP 44346, 31028 Toulouse cedex 4, France}

\author{Kazuki Tokuda}
\affil{Department of Physical Science, Graduate School of Science, Osaka Prefecture University, 1-1 Gakuen-cho, Naka-ku, Sakai, Osaka 599-8531, Japan}
\affil{National Astronomical Observatory of Japan, 2-21-1 Osawa, Mitaka, Tokyo, 181-8588, Japan}

\author{R\'emy Indebetouw}
\affil{Department of Astronomy, University of Virginia, P.O. Box 400325, Charlottesville, VA 22904, USA}
\affil{National Radio Astronomy Observatory, 520 Edgemont Road Charlottesville, VA 22903, USA}

\author{Jean-Philippe Bernard}
\affil{Universit\'{e} de Toulouse, UPS-OMP, 31028 Toulouse, France}
\affil{CNRS, IRAP, Av. du Colonel Roche BP 44346, 31028 Toulouse cedex 4, France}

\author{Toshikazu Onishi}
\affil{Department of Physical Science, Graduate School of Science, Osaka Prefecture University, 1-1 Gakuen-cho, Naka-ku, Sakai, Osaka 599-8531, Japan}

\author{Evan Wojciechowski}
\affil{Department of Astronomy, University of Illinois, 
Urbana, IL 61801, USA}

\author{Jeffrey B. Bandurski}
\affil{Department of Astronomy, University of Illinois, 
Urbana, IL 61801, USA}

\author{Akiko Kawamura}
\affil{National Astronomical Observatory of Japan, 2-21-1 Osawa, Mitaka, Tokyo, 181-8588, Japan}

\author{Julia Roman-Duval}
\affil{Space Telescope Science Institute, 3700 San Martin Drive, Baltimore, MD 21218, USA}

\author{Yixian Cao}
\affil{Department of Astronomy, University of Illinois, 
Urbana, IL 61801, USA}

\author[0000-0002-3925-9365]{C.-H. Rosie Chen}
\affil{Max-Planck-Institut f\"ur Radioastronomie, Auf dem H\"ugel 69, D-53121 Bonn, Germany}

\author{You-hua Chu}
\affil{Department of Astronomy, University of Illinois, 
Urbana, IL 61801, USA}
\affil{Academia Sinica Institute of Astronomy and Astrophysics, Taipei 10617, Taiwan, Republic of China}

\author[0000-0001-6096-0757]{Chaoyue Cui}
\affil{Department of Astronomy, University of Illinois, 
Urbana, IL 61801, USA}

\author{Yasuo Fukui}
\affil{Department of Physics, Nagoya University, Chikusa-ku, Nagoya 464-8602, Japan}

\author{Ludovic Montier}
\affil{Universit\'{e} de Toulouse, UPS-OMP, 31028 Toulouse, France}
\affil{CNRS, IRAP, Av. du Colonel Roche BP 44346, 31028 Toulouse cedex 4, France}

\author{Erik Muller}
\affil{National Astronomical Observatory of Japan, 2-21-1 Osawa, Mitaka, Tokyo, 181-8588, Japan}

\author{Juergen Ott}
\affil{National Radio Astronomy Observatory, 1003 Lopezville Rd, Socorro, NM 87801, USA}

\author{Deborah Paradis}
\affil{Universit\'{e} de Toulouse, UPS-OMP, 31028 Toulouse, France}
\affil{CNRS, IRAP, Av. du Colonel Roche BP 44346, 31028 Toulouse cedex 4, France}

\author[0000-0001-8898-2800]{Jorge L. Pineda}
\affil{Jet Propulsion Laboratory, California Institute of Technology, 4800 Oak Grove Drive, Pasadena, CA 91109, USA}

\author[0000-0002-5204-2259]{Erik Rosolowsky}
\affil{Department of Physics, University of Alberta, Edmonton, AB, T6G 2E1, Canada}

\author[0000-0003-2248-6032]{Marta Sewi{\l}o}
\affil{NASA Goddard Space Flight Center, 8800 Greenbelt Rd, Greenbelt, MD 20771, USA}

\accepted{10 October 2017 by ApJ}

\begin{abstract}
We present high-resolution (sub-parsec) observations of a giant molecular cloud in the nearest star-forming galaxy, the Large Magellanic Cloud.  ALMA Band 6 observations trace the bulk of the molecular gas in $^{12}$CO(2--1) and high column density regions in $^{13}$CO(2--1).  Our target is a quiescent cloud (PGCC G282.98$-$32.40, which we refer to as the ``Planck cold cloud'' or PCC) in the southern outskirts of the galaxy where star-formation activity is very low and largely confined to one location.  We decompose the cloud into structures using a dendrogram and apply an identical analysis to matched-resolution cubes of the 30 Doradus molecular cloud (located near intense star formation) for comparison.  Structures in the PCC exhibit roughly 10 times lower surface density and 5 times lower velocity dispersion than comparably sized structures in 30 Dor, underscoring the non-universality of molecular cloud properties.  In both clouds, structures with relatively higher surface density lie closer to simple virial equilibrium, whereas lower surface density structures tend to exhibit super-virial line widths.  In the PCC, relatively high line widths are found in the vicinity of an infrared source whose properties are consistent with a luminous young stellar object.  More generally, we find that the smallest resolved structures (``leaves'') of the dendrogram span close to the full range of line widths observed across all scales.  
As a result, while the bulk of the kinetic energy is found on the largest scales, the small-scale energetics tend to be dominated by only a few structures, leading to substantial scatter in observed size-linewidth relationships.
\end{abstract}


\section{Introduction}

Molecular clouds are highly turbulent (i.e., their observed line widths are suprathermal), and they exhibit a correlation between size and line width, approximately of the form $\sigma_v \propto R^{1/2}$ (hereafter the $R$-$\sigma_v$ relation), that is reminiscent of a Kolmogorov-type cascade.  Although the $R$-$\sigma_v$ relation has been studied for decades \citep{Larson:81,Solomon:87}, interpretation has been hampered by a number of observational biases.  Molecular cloud sizes are dependent on the observational resolution and sensitivity, and observations of molecular lines are sensitive to a narrow range of densities and could miss very dense or very diffuse structures.  Line opacity can affect the observed line width and lead to blending of multiple components \citep{Hacar:16b}, and comparison of clouds at different distances must take into account beam dilution effects \citep[e.g.,][]{Goodman:98}. Despite these challenges, it has been customary to interpret the observed $R$-$\sigma_v$ relation in terms of supersonic turbulence driven by kinetic energy injection \citep{Falgarone:94,Brunt:02,Kritsuk:13}.

A key result which has stimulated a resurgence of interest in the $R$-$\sigma_v$ relation was presented by \citet{Heyer:09}, who demonstrated using data from the BU-FCRAO Galactic Ring Survey (GRS) of $^{13}$CO(1--0) emission that the normalization of the relation ($v_0 = \sigma_v/R^{1/2}$) exhibits a linear correlation with mass surface density ($\Sigma=M/\pi R^2$) across more than an order of magnitude in $\Sigma$.  The data are consistent with a model of clouds in virial equilibrium, except that the observed normalization $v_0$ is about a factor of 2 too large.  Subsequent work by \citet{Field:11} has suggested that the larger than expected $v_0$ may result from external pressure confinement, although to a lesser extent than has been inferred for clouds in the outer Galaxy \citep{Heyer:01} or near the Galactic Center \citep{Oka:01}.  On the other hand, \citet{Ballesteros:11a} have interpreted the \citet{Heyer:09} result in terms of gravitational collapse near free-fall, which differs from the virial equilibrium prediction by a factor of $\sqrt{2}$ in $v_0$, and thus is roughly consistent with the GRS data.  Although these interpretations differ markedly, they share a common theme of departing from the standard view of molecular clouds being in a state of balance between gravity and random turbulent motions.

The present work aims at improving our empirical knowledge of molecular cloud structure and turbulence by comparing two molecular clouds in the nearest star-forming galaxy, the Large Magellanic Cloud.
The proximity ($d \approx 50$ kpc, so $1\arcsec \approx 0.24$ pc) and face-on aspect of the Large Magellanic Cloud provides the opportunity to study a population of well-separated molecular clouds at a common distance.  The population of molecular clouds in the LMC has been previously surveyed by the NANTEN \citep{Fukui:08} and MAGMA \citep{Wong:11} CO surveys.  The 40-pc resolution NANTEN survey identified 272 giant molecular clouds (GMCs) with masses $\gtrsim 2 \times 10^4\,M_\odot$.  The MAGMA survey mapped a flux-selected sample of $\sim$160 of these clouds at higher (11 pc) resolution, resolving them into 450 regions of contiguous emission (``islands'').  \citet{Hughes:10} and \citet{Wong:11} analyzed the physical properties of the MAGMA GMCs, finding that the $R$-$\sigma_v$ relation exhibits substantial scatter while displaying a systematic offset toward smaller line widths compared to the Milky Way clouds studied by \citet{Solomon:87}.

On scales of $\lesssim$10 pc, molecular line observations probe the substructure of molecular clouds, producing a ``Type-4'' (single-cloud, single-tracer) $R$-$\sigma_v$ relation in the terminology of \citet{Goodman:98}.  With the resolution provided by the Atacama Large Millimeter/submillimeter Array (ALMA), it is now possible to conduct such observations in the LMC.  Here we focus on a particularly quiescent cloud in the LMC which has been designated as Planck Galactic Cold Clump (PGCC) G282.98$-$32.40 \citep{Planck:XXVIII}.  Based on estimated dust temperatures inferred from Planck data at 4\arcmin\ resolution, it is very cold ($T_d \lesssim 15$ K) and lacking in active massive star formation.  Although the cloud falls outside the region surveyed by NANTEN, its size and mass are consistent with GMCs detected by NANTEN, and it was also detected in the {\it Planck} integrated CO(1--0) map.  We subsequently mapped the cloud in CO(1--0) with the Mopra antenna of the Australia Telescope National Facility (see \S\ref{sec:mopra}), confirming the presence of strong CO emission, before obtaining high-resolution observations with the ALMA.  For convenience we hereafter refer to the target as the ``Planck cold cloud'' or ``PCC,'' although we emphasize that it is just one of several such clouds identified in the periphery of the LMC \citep{Planck:XXVIII}.

This paper is organized as follows.  We present the observational data for the PCC in \S\ref{sec:obs}, including large-scale view of the cloud from Mopra (\S\ref{sec:mopra}) and the high-resolution mapping with ALMA (\S\ref{sec:alma}).  We present total flux information in \S\ref{sec:ratios}, confirm the quiescent nature of the PCC by analyzing available infrared imaging in \S\ref{sec:tdust}, and undertake a structural decomposition of the cloud using dendrograms in \S\ref{sec:dendro} and \S\ref{sec:corrs}.  We use a local thermodynamic equilibrium (LTE) analysis to estimate the column densities of the substructures in \S\ref{sec:lte}.  Both the decomposition and LTE analysis are also performed on a previous ALMA data set of the 30 Doradus cloud \citep{Indebetouw:13}, where star formation is much more active, to provide a point of comparison.  We discuss our results in \S\ref{sec:disc} and summarize our conclusions in \S\ref{sec:conc}.

\section{Observations and Data Reduction}\label{sec:obs}

\subsection{Mopra CO(1--0) Data}\label{sec:mopra}

The MAGMA survey presented by \citet{Wong:11} includes observations made in 2005--2010 of a subset of CO clouds identified by the NANTEN survey \citep{Fukui:08}.  Several additional projects undertaken with Mopra in 2012--3 (M624, PI: Roman-Duval; M625, PI: Bernard; M633, PI: Wong; M1022, PI: Hughes) expanded the MAGMA areal coverage by $\sim$20\% to include additional fainter clouds, as well as clouds falling outside the areal coverage of the NANTEN survey.  The region of particular interest for this paper was observed as part of project M625 in 2013.  All of the additional regions observed in 2012--3 have been incorporated into the Third MAGMA Data Release\footnote{http://mmwave.astro.illinois.edu/magma/DR3/} (DR3), which we document for completeness here.

\begin{figure*}[tbh]
\includegraphics[height=3.4in]{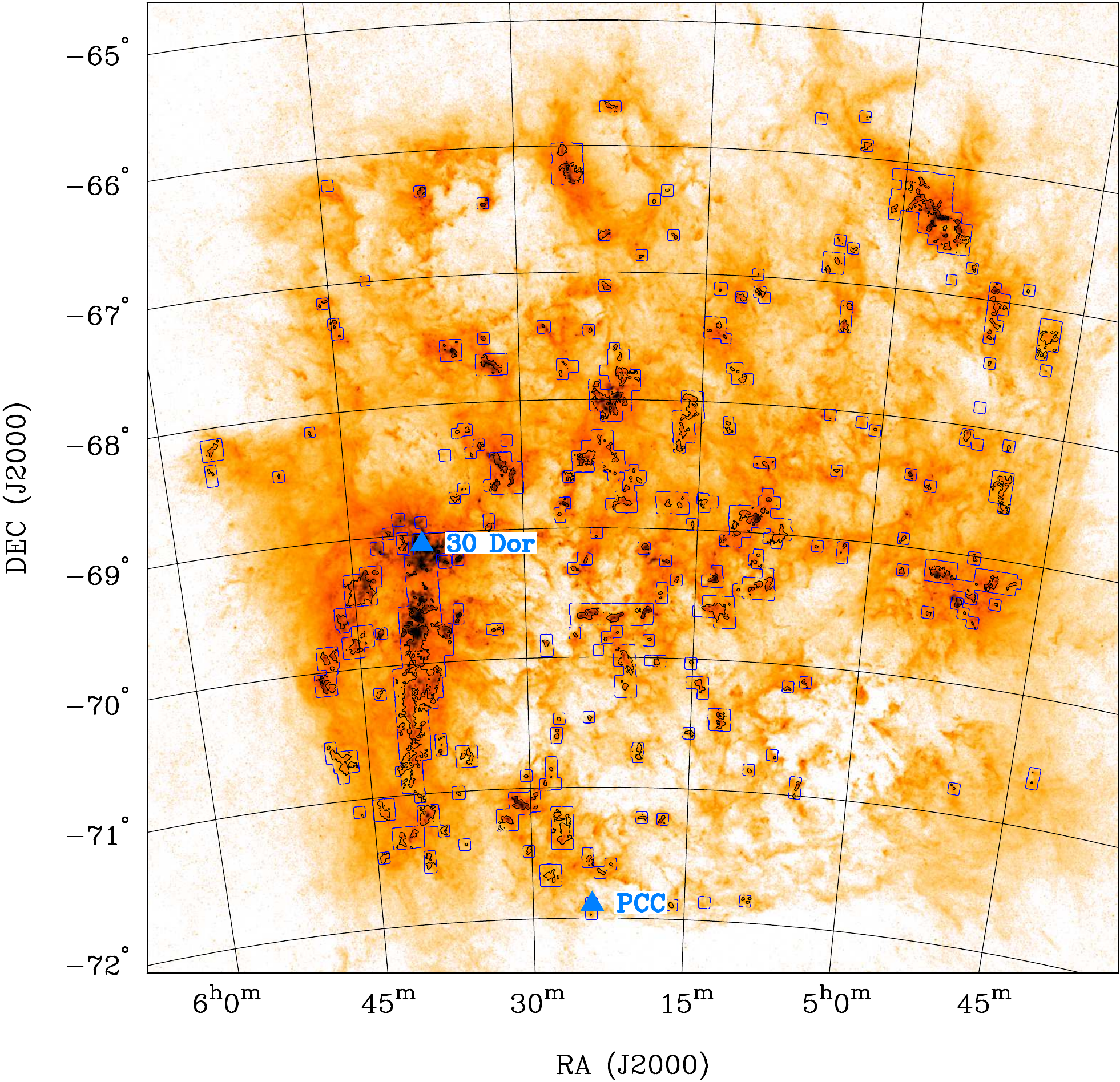}\hfill
\includegraphics[height=3.4in]{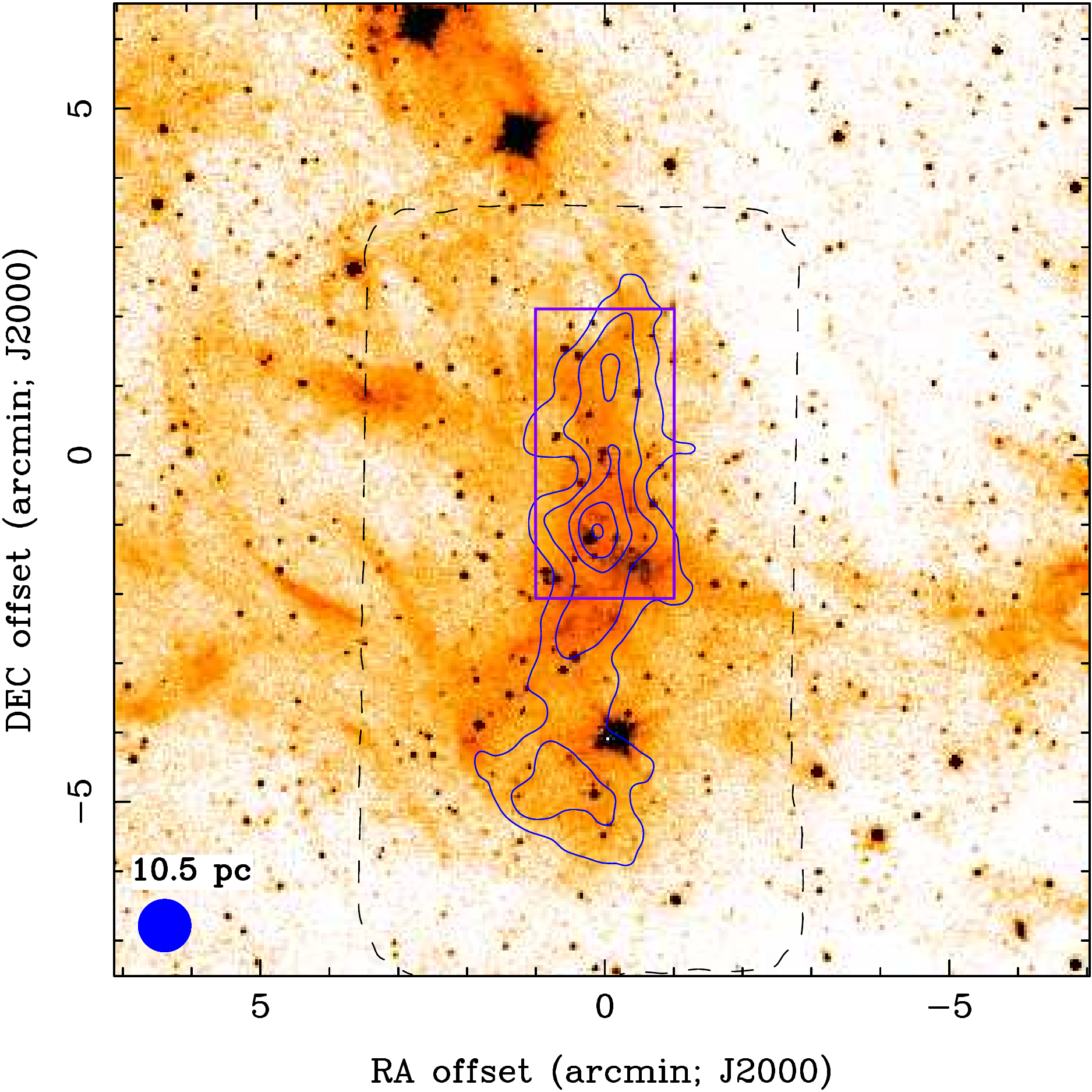}
\caption{\small{{\it Left}: MAGMA DR3 CO(1--0) areal coverage and intensity map overlaid on the HERITAGE 350 $\mu$m image.  Rectangular blue contours outline the regions mapped in CO, while the black contour shows the CO intensity at the 2.5 K \kms\ level.  The highlighted cloud at the southern edge of the galaxy is the focus of this paper.  {\it Right}: Zoomed-in view of the PCC cloud, centered on $\alpha_{2000}$=5$^{\rm h}$24$^{\rm m}$09\farcs2 and $\delta_{2000}$=$-71$\arcdeg53\arcmin37\arcsec, with MAGMA contours at levels of 1, 2, ..., 5 K \kms\ overlaid on the SAGE 8 $\mu$m image.  
The 45\arcsec\ MAGMA resolution is indicated by the circle in the lower left corner.
The magenta rectangle outlines the region observed with ALMA, while the dashed black contour outlines the region observed with Mopra.
}}
\label{fig:magma}
\end{figure*}

We employed the same mapping approach as with MAGMA, scanning 5\arcmin\ $\times$ 5\arcmin\ square regions to reach a map sensitivity of $\sigma(T_{\rm mb}) \approx 0.3$ K per 0.526 \kms\ channel by making two passes, one in right ascension and the other in declination.  A single pass required about 75 minutes to complete and was followed by a pointing correction on the SiO maser R Doradus (typical pointing corrections were about 5\arcsec).  The CO spectra were placed on a main-beam brightness temperature scale ($T_{\rm mb}$) assuming an ``extended beam'' efficiency of 0.4 based on daily observations of Orion KL referenced to the measurements of \citet{Ladd:05}.  Spectra were subsequently gridded into a data cube using the ATNF package {\sc gridzilla}.  The resulting maps possess a Gaussian beam of 45\arcsec\ FWHM which is oversampled with a pixel scale of 15\arcsec.  The CO(1--0) intensity map from MAGMA DR3 is presented in Figure~\ref{fig:magma}(a), overlaid on the HERITAGE 350 $\mu$m image \citep{Meixner:13}.

As shown in Figure~\ref{fig:magma}(b), where the MAGMA CO contours are overlaid on the SAGE 8 $\mu$m map \citep{Meixner:06}, the PCC cloud was covered by two adjacent 5\arcmin\ square maps.  To generate an integrated CO image, the cube was first smoothed spatially, to reach a resolution of 67\farcs5, and spectrally, by a convolving with a Gaussian with a FWHM of 3 channels.  Then a mask was generated by expanding from the 5$\sigma$ contour of the smoothed cube out to the 3$\sigma$ contour. (Both the initial and expanded mask are required to be at least 2 channels wide at every spatial pixel.) This mask was then applied to the original data cube before summing along the velocity axis.  The total CO flux measured within the mask was 1040 Jy km s$^{-1}$, which translates to a molecular gas mass (including helium) of $2.7 \times 10^4$ \Msol\ using a standard $\alpha_{\rm CO} = 4.3$ \Msol\ pc$^{-2}$ (K km s$^{-1}$)$^{-1}$ [equivalent to $X_{\rm CO} = 2 \times 10^{20}$ cm$^{-2}$ (K km s$^{-1}$)$^{-1}$] \citep{Bolatto:13a}.  We note that values of $\alpha_{\rm CO}$ roughly a factor of 2 larger than the standard value have been derived for the LMC using virial or dust-based mass estimators \citep{Hughes:10,Leroy:11,Jameson:16}, which would require that the mass quoted here be scaled up by a similar factor.

\begin{figure*}[tbh]
\includegraphics[scale=0.75]{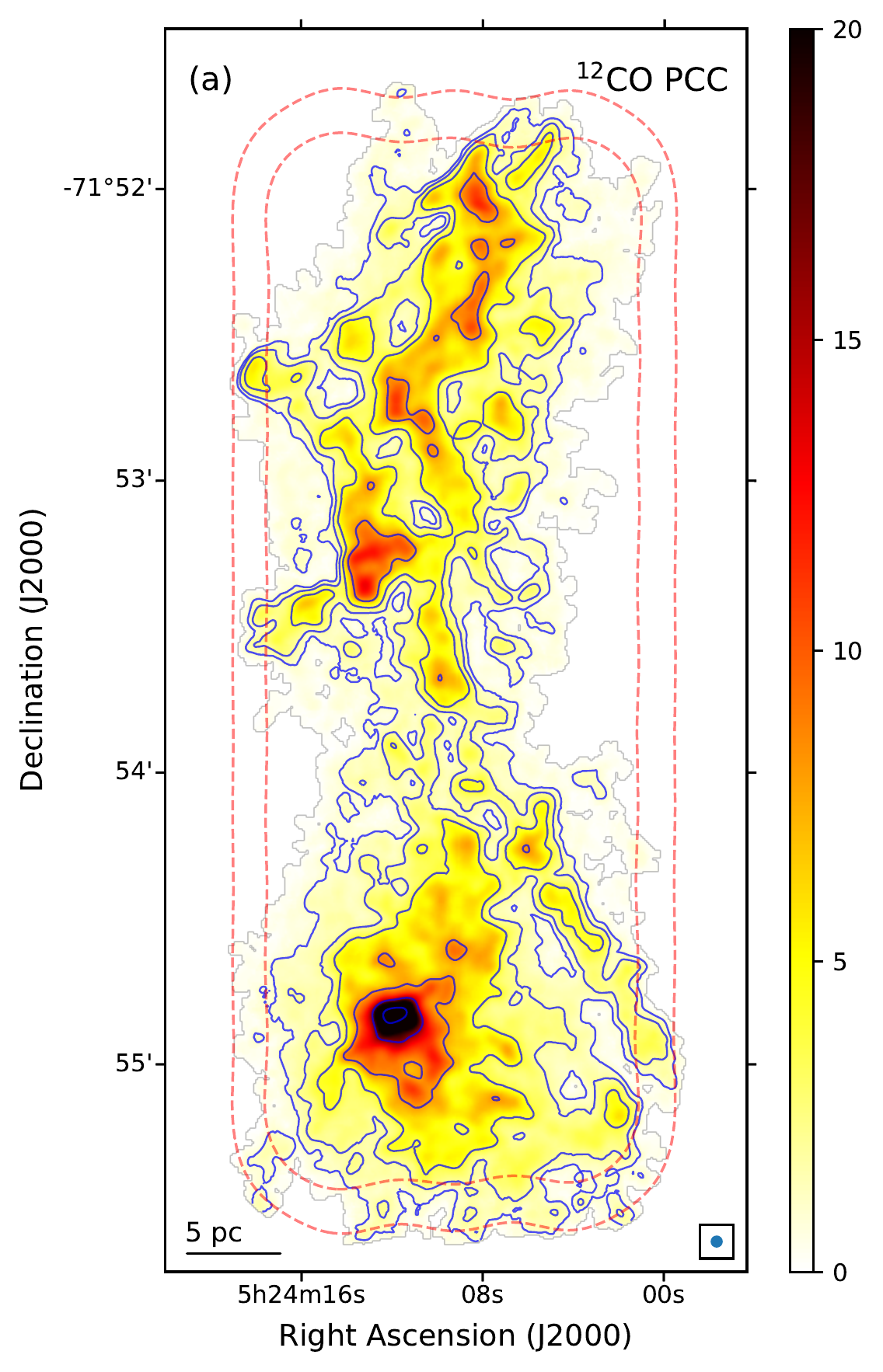}\hfill
\includegraphics[scale=0.75]{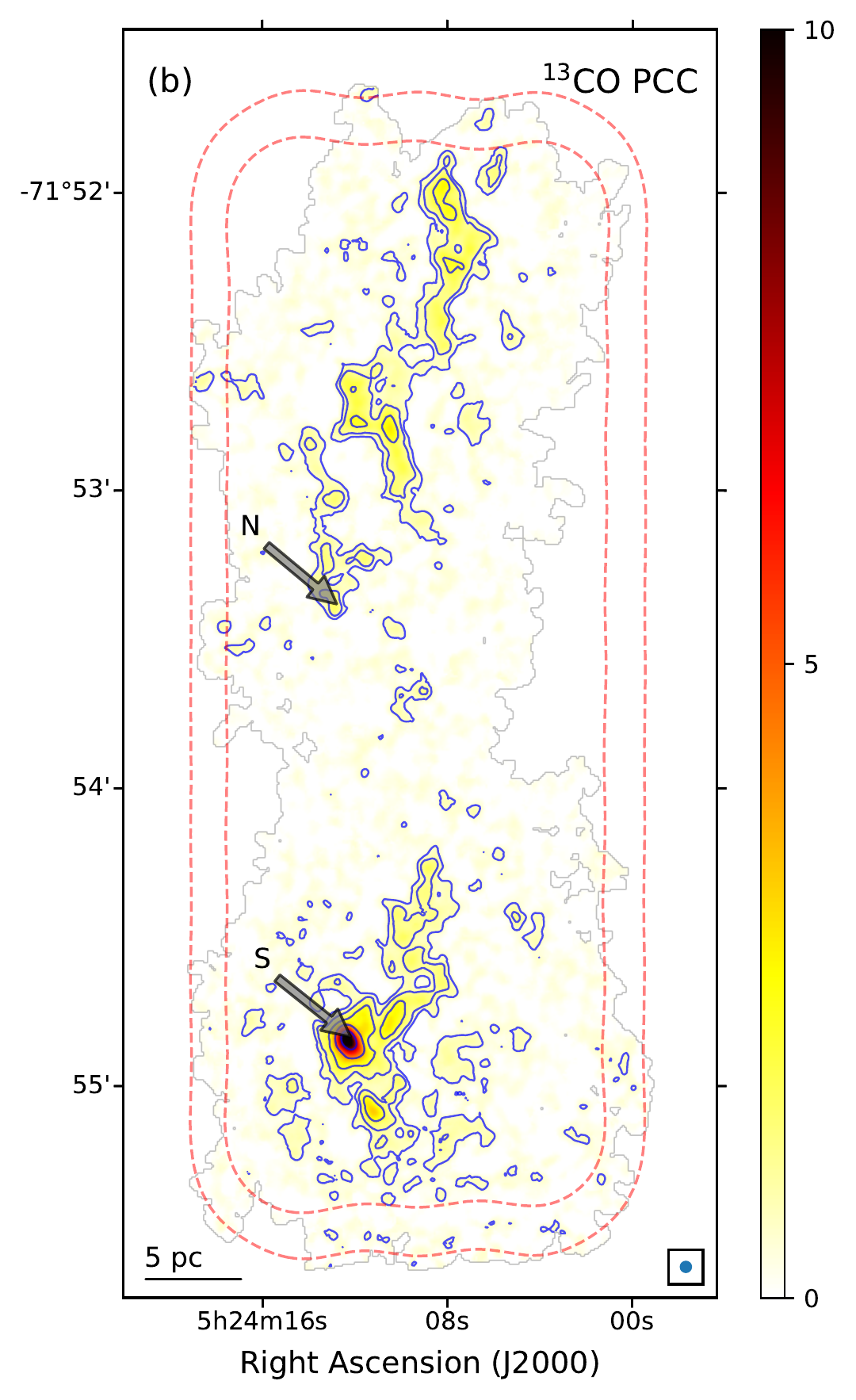}\\[2ex]
\includegraphics[scale=0.72]{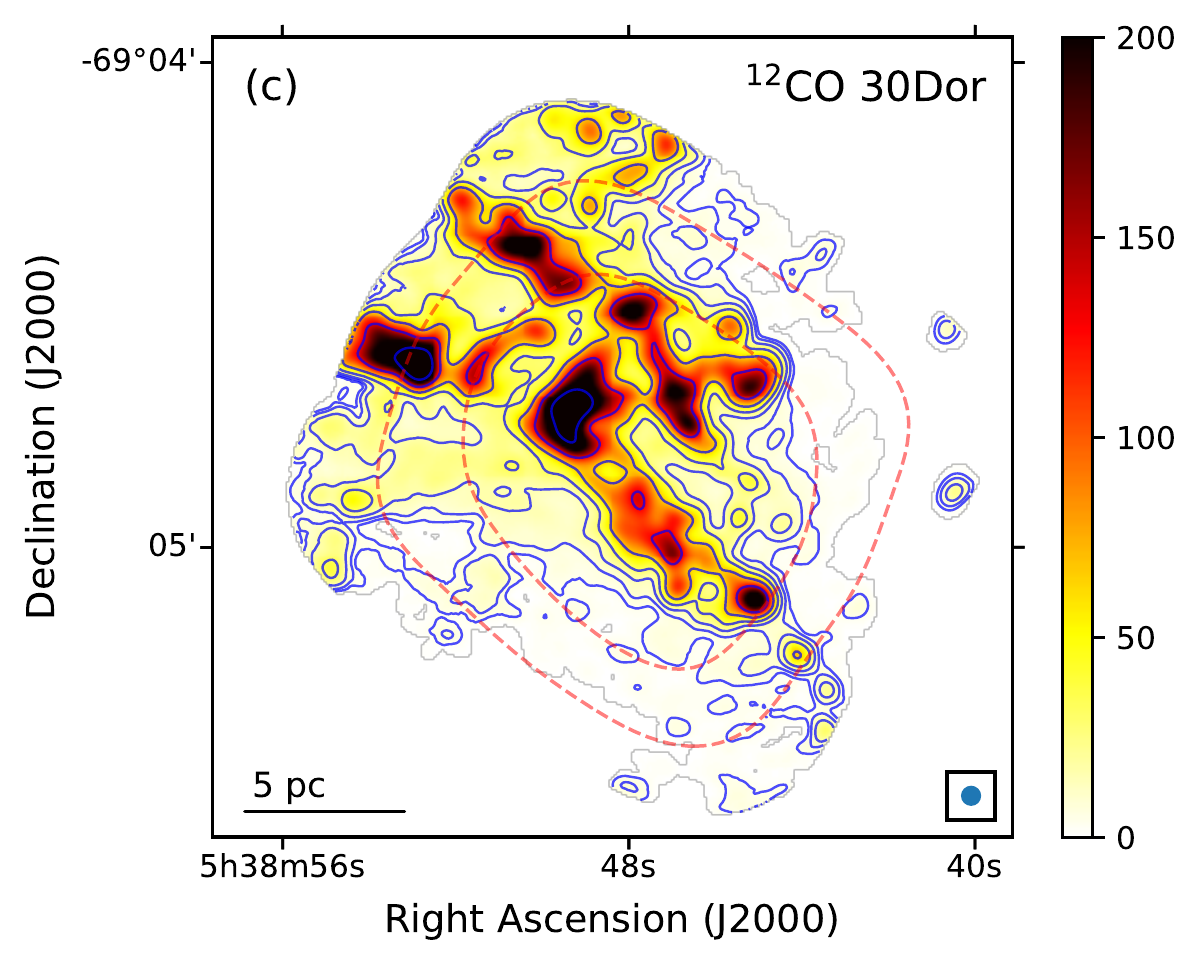}\hfill
\includegraphics[scale=0.72]{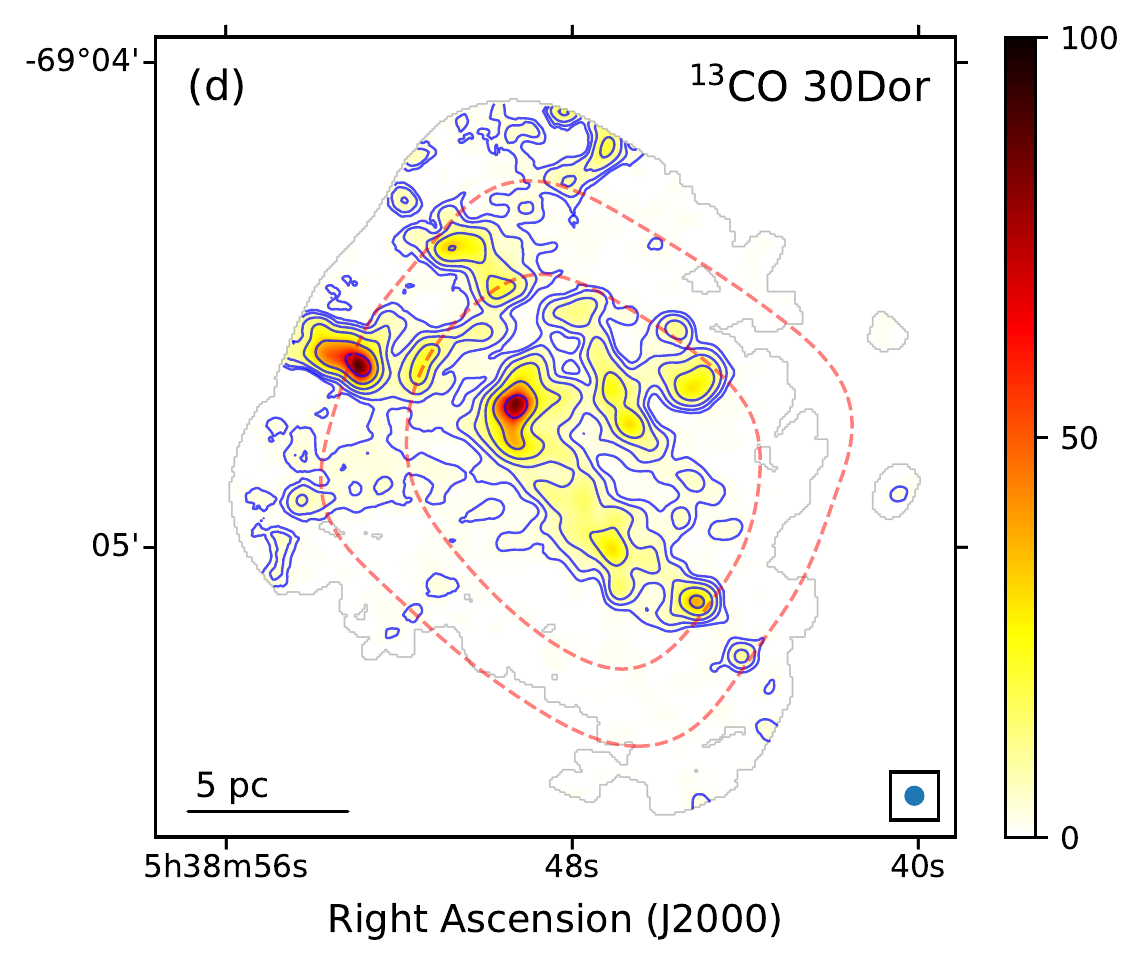}
\caption{\small{CO(2--1) and $^{13}$CO(2--1) intensity maps for the PCC ({\it upper panels}) and 30 Dor region ({\it lower panels}), at a common resolution of 2\farcs5, obtained by integrating the ALMA cubes over the region of the dilated CO mask.  Contour levels are $2^n$ K \kms, where $n$=0, 1, ..., 6 for panel (a), $n$=$-1$, 0, ..., 5 for panel (b), $n$=2, 3, ..., 8 for panel (c), and $n$=1, 2, ..., 7 for panel (d).  Color scales are also given in K \kms\ units.  Physical scales and beam sizes are indicated on the bottom corners of each panel.  The red dashed contours indicate 50\% and 80\% of the peak sensitivity in the ALMA mosaic.  Arrows in panel (b) indicate the positions for which spectra are displayed in Fig.~\ref{fig:spectra}.
}}
\label{fig:mom0}
\end{figure*}

\subsection{ALMA CO(2--1) and $^{13}$CO(2--1) Data}\label{sec:alma}

ALMA Cycle 2 observations were obtained in 2014 and 2015 under project code 2013.1.00832.S (PI: Wong).  The observations cover a region of 220\arcsec\ $\times$ 80\arcsec\ elongated in the north-south direction, covering the northern half of the PCC cloud. [Figure~\ref{fig:magma} {\it (right)}, Table~\ref{tab:regs}].  The correlator was set to simultaneously observe the $^{12}$CO(2--1), $^{13}$CO(2--1), and C$^{18}$O(2--1) lines with a velocity resolution of 0.1 \kms, along with two continuum bands at 218 and 232 GHz.  Observations were performed separately using the 12m, 7m (ACA), and total power (TP) arrays, and calibrated data sets were delivered via the ALMA pipeline.

Because the pipeline-reduced 12m data showed obvious negative artifacts, we re-imaged the calibrated visibilities in CASA.  This was performed initially using automatically generated CLEAN boxes (as described in the M100 Band 3 tutorial of the online CASA Guides) followed by a final interactive CLEAN.  The resulting 12m images were combined with the corresponding TP images using the CASA task {\tt feather}.  In this study we have not incorporated the ACA data, given that the ACA cubes have limited sensitivity ($1\sigma$ noise of 200 \mjybm\ in a 0.2 \kms\ channel, twice larger than the 12m data convolved to the same resolution).  We compared the flux distribution of the 12m+TP cube with two methods of incorporating the ACA data (in the visibility plane or by feathering) and found the fluxes to differ by $\lesssim$10\%.

Integrated intensity images of the CO(2--1) and $^{13}$CO(2--1) lines are shown in Figure~\ref{fig:mom0}.  As with the MAGMA data, a signal mask was applied to the cube prior to summing in velocity.  The mask was obtained by starting at the 3.5$\sigma$ contour and expanding to the 2$\sigma$ contour, then padding the mask by 1 pixel in all directions.  The 1mm continuum emission was not detected at an RMS noise level of 0.21 mJy beam$^{-1}$ for a 1\farcs58 $\times$ 1\farcs02 beam, while the C$^{18}$O(2--1) line was only marginally detected at the location of the southern peak.  For the remainder of the paper we focus on the CO and $^{13}$CO data.

For comparison purposes, we have also analyzed ALMA data for an additional cloud, 30Dor-10 \citep{Indebetouw:13}, observed in Cycle 0 under project code 2011.0.00471.S (PI: Indebetouw).  
Prior to further analysis, we convolve the data for both clouds to achieve a circular Gaussian beam of FWHM 2\farcs5, corresponding to 0.6 pc.
The characteristics of the maps before and after resolution matching are summarized in Table~\ref{tab:regs}.

\begin{deluxetable*}{lccccccccc}
\tablehead{
\colhead{Region} & \colhead{Ref.\ R.A.} & \colhead{Ref.\ Dec.} & \colhead{Mosaic Size} & \colhead{Beam Size} & \colhead{$\Delta v_{\rm ch}$} & \colhead{$T_{\rm rms,12}$\tablenotemark{a}} & \colhead{$T_{\rm peak,12}$\tablenotemark{b}} & \colhead{$T_{\rm rms,13}$} & \colhead{$T_{\rm peak,13}$}\\
\colhead{} & \colhead{(J2000)} & \colhead{(J2000)} & \colhead{$(\arcsec \times \arcsec)$} & \colhead{$(\arcsec \times \arcsec)$} & \colhead{(km s$^{-1}$)}
& \colhead{(K)} & \colhead{(K)} & \colhead{(K)} & \colhead{(K)}}
\tablecaption{Summary of Map Parameters\label{tab:regs}}
\startdata
PCC & 5$^{\rm h}$24$^{\rm m}$09\fs2 & $-71$\degr 53\arcmin 37\arcsec & 80 $\times$ 220 & 1.81 $\times$ 1.24 & 0.2 & 0.43 & 14.7 & 0.46 & \phn7.9\\
 &  &  &  & 2.50 $\times$ 2.50 & 0.2 & 0.19 & 12.0 & 0.21 & \phn5.3\\
30 Dor & 5$^{\rm h}$38$^{\rm m}$47\fs0 & $-69$\degr 04\arcmin 36\arcsec & 50 $\times$ \phn50 & 2.47 $\times$ 1.59 & 0.5 & 0.12 & 60.3 & 0.13 & 23.9\\
 &  &  &  & 2.50 $\times$ 2.50 & 0.5 & 0.07 & 51.5 & 0.08 & 19.6\\
\enddata
\tablenotetext{a}{RMS noise in a channel map of width $\Delta v_{\rm ch}$ in the $^{12}$CO(2--1) cube.}
\tablenotetext{b}{Peak brightness temperature of the $^{12}$CO(2--1) cube.}
\end{deluxetable*}

\section{Results}

\subsection{Fluxes and Line Ratios From Total Power Data}\label{sec:ratios}

Since ALMA provides total power (TP) data for the CO(2--1) and $^{13}$CO(2--1) lines (at 28\arcsec\ resolution), which can be compared with the MAGMA CO(1--0) map at 45\arcsec\ resolution, we have derived integrated line ratios for the PCC cloud for comparison with previously published work.
Integrating over the observed ALMA field, we measure a CO(2--1) flux of 1691$\pm$10 Jy \kms\ and a $^{13}$CO(2--1) flux of 125$\pm$15 Jy \kms, implying a CO(2--1)/$^{13}$CO(2--1) flux ratio of 13.5$\pm$1.6.  
The quoted uncertainties reflect half the difference in flux measured with and without a signal mask; the formal uncertainties due to map noise are much smaller ($\sim$1\%) because of the very high signal-to-noise ratio (up to 30--70 in each 0.2 \kms\ channel) of the TP data.  Uncertainties in the absolute flux scale are expected to be $<$10\%\footnote{based on the ALMA Cycle 2 Proposer's Guide} and are not included in these uncertainties---since the lines are observed simultaneously, errors in the flux scale should not affect the line ratio.
Convolving the TP data to the 45\arcsec\ resolution of the MAGMA data yields a CO(2--1)/CO(1--0) ratio of $\sim$2.5$\pm$0.5 in flux (Jy) units or $\sim$0.6$\pm$0.1 in brightness temperature units, assuming calibration uncertainties of 20\% and 10\% for the MAGMA and TP data respectively.  The integrated TP spectrum has a linewidth of $\sigma_v = 1.00\pm.01$ \kms.

\citet{Wang:09} measured a CO(2--1)/$^{13}$CO(2--1) ratio of 5.7 toward N113 (using the SEST with a 24\arcsec\ beam), indicating higher average line opacities in N113 compared to the PCC, consistent with the much stronger line fluxes from N113. 
\citet{Sorai:01} measured CO(2--1)/CO(1--0) ranging from 0.5 to 1.3 for 34 CO(1--0) peaks across the LMC at 9\arcmin\ resolution, with a luminosity-weighted average ratio of 0.9.  Although their average value is larger than the value we obtain for the PCC (0.6), \citet{Sorai:01} report that clouds in the outer parts of the LMC exhibit lower CO(2--1)/CO(1--0) ratios than clouds in the inner LMC, consistent with lower densities and temperatures for the outer clouds.

\subsection{Dust Temperature Validation}\label{sec:tdust}

The PCC cloud was originally identified as a very cold region in a map of the LMC's dust temperature that was constructed from Planck data with a resolution of $\sim5$\arcmin. This is considerably larger than both the angular resolution and the field-of-view (FoV) of the ALMA mosaic presented here. To cross-check our assumption that the PCC cloud is one of the coldest molecular gas regions in the LMC, we re-estimated the average dust temperature of the PCC, and compared it to the 30~Doradus molecular cloud and other CO-emitting regions identified in the MAGMA survey. For each region, we estimated the dust temperature using a simple modified blackbody (MBB) model for the spectral energy distribution (SED) of the thermal dust emission at wavelengths between 100 and 500\,$\mu$m. The input SED was constructed using the MIPS, PACS and SPIRE data delivered by the Spitzer SAGE and Herschel HERITAGE programs \citep{Meixner:06,Meixner:13}. All bands were convolved to a common resolution of 1\arcmin. For each region, the MBB fit is performed using an iterative procedure based on the Levenberg-Marquardt method, as implemented in the {DUSTEM} software package \citep{Compiegne:11}.\footnote{Available from {\tt http://dustemwrap.irap.omp.eu/}.} The fitted parameters of the MBB model are the dust temperature $T_{d}$, the spectral index $\beta$ and the normalization of the dust SED. We constrain $T_{d}$ and $\beta$ to lie in the intervals [5, 55]\,K and [0.8, 2.8] respectively. We assume a 10\% total uncertainty in the average flux densities at all bands; for simplicity, we do not impose any correlation between bands in these uncertainties. Although neglecting correlated uncertainties can lead to significant mis-estimates of dust parameters \citep{Gordon:14}, we stress that we are primarily interested here in the {\it relative} temperatures of GMCs.

We construct a single SED for each CO-emitting region using median values of the flux density across the region, i.e.\ we do not probe spatial variations of the dust temperature within clouds. To isolate the emission from a target molecular region, a local background surrounding the region was defined and the median surface brightness at each band within this background was subtracted from the median surface brightness measured within the target region. For the MAGMA clouds, we use the `islands' assignment mask presented by \citet{Wong:11} to define the target CO-emitting regions. For the PCC and 30~Doradus regions, we use two different masks: the field of view of the ALMA mosaics (mask 1), and the more extended region of CO emission identified by MAGMA enclosing the ALMA field (mask 2). We note that some difference in the dust temperatures estimated using the different masks is to be expected since the ALMA observations focus on a subregion of the molecular cloud identified by MAGMA. For all clouds, the background regions that we use have roughly twice the projected spatial extent as the target molecular regions. Sightlines with a significant detection of emission in the MAGMA $^{12}$CO(1--0) cube are excluded from our defined background regions. 

\begin{figure}[tbh]
\begin{center}
\includegraphics[width=0.45\textwidth, viewport=70 18 690 675,clip=true]{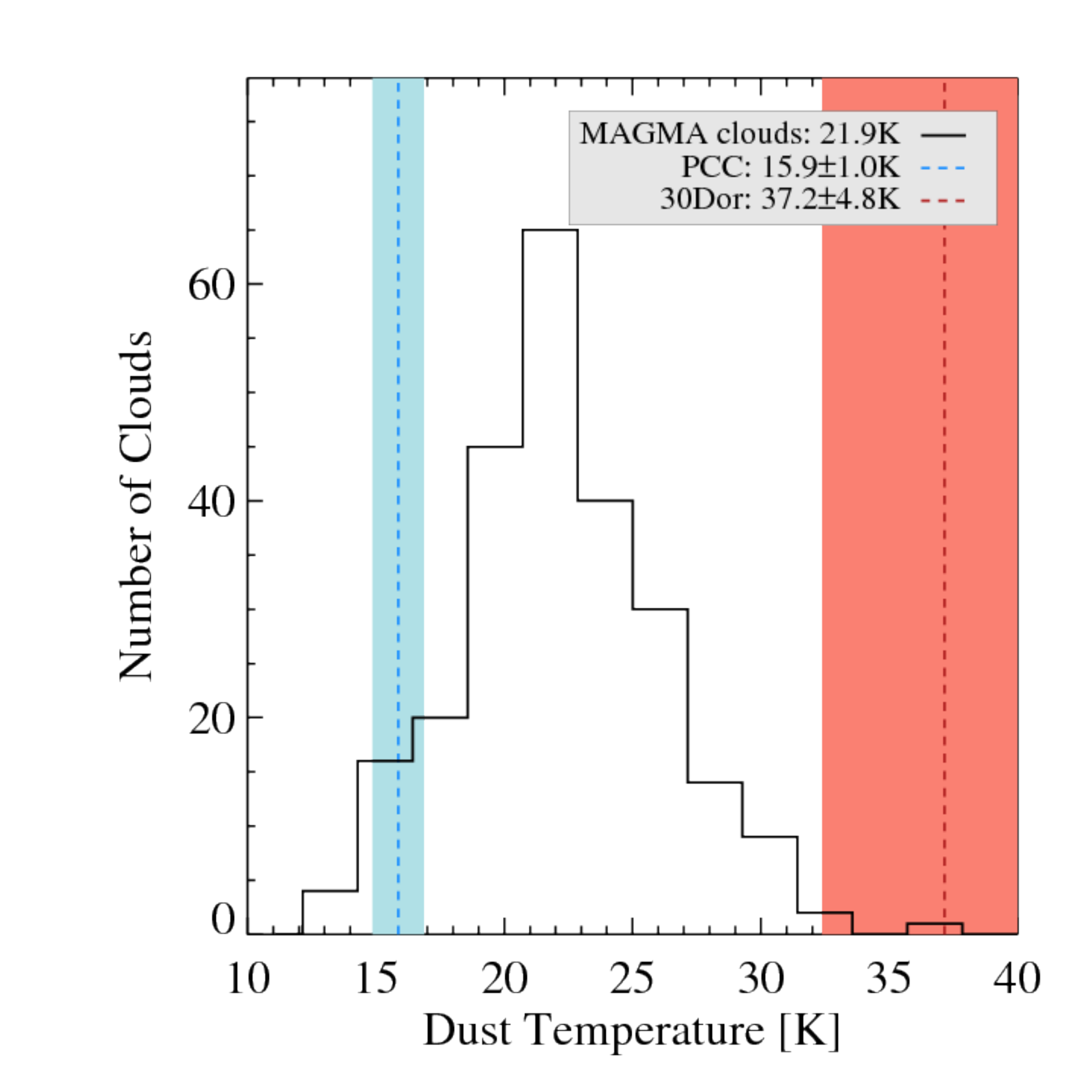}
\end{center}
\caption{Distribution of dust temperatures for CO-emitting regions identified in the MAGMA survey. The median dust temperature of the MAGMA clouds is 21.9\,K. The average dust temperatures of the PCC and 30~Doradus cloud regions are indicated with blue and red vertical lines, with colored shading representing the uncertainty in their dust temperature estimates.  \label{fig:tdust_histo}}
\end{figure}

Figure~\ref{fig:tdust_histo} presents the distribution of dust temperatures that we obtain for the CO-emitting `islands' identified in the MAGMA survey. The dust temperatures that we infer for the PCC and 30~Doradus molecular clouds are overplotted as blue and red lines, with shaded regions used to indicate the uncertainty. For consistency with the data points in the histogram, we overplot the average dust temperatures of PCC and 30~Doradus within mask 2. Figure~\ref{fig:tdust_histo} clearly shows that the PCC and 30~Doradus lie at the opposite extremes of dust temperatures observed for molecular clouds in the LMC. The PCC and 30~Dor clouds have dust temperatures of $15.9\pm1.0$\,K and $37\pm5$\,K respectively, while the bulk of the LMC cloud population (defined using the 10$^{\rm th}$ and 90$^{\rm th}$ percentiles) has a dust temperature between 17 and 27\,K. The dust temperatures inferred for the PCC and 30~Dor regions within the more restricted field of view of the ALMA observations (mask 1) are $17.4\pm1.2$\,K and $30.1\pm3.6$\,K.  Thus we confirm the very cold nature of the PCC.

\subsection{Structural Decomposition}\label{sec:dendro}

\begin{figure*}[tbh]
\begin{minipage}{\textwidth}
  \centering
  \raisebox{-0.5\height}{\includegraphics[width=2.2in,viewport=5 5 263 478,clip=true]{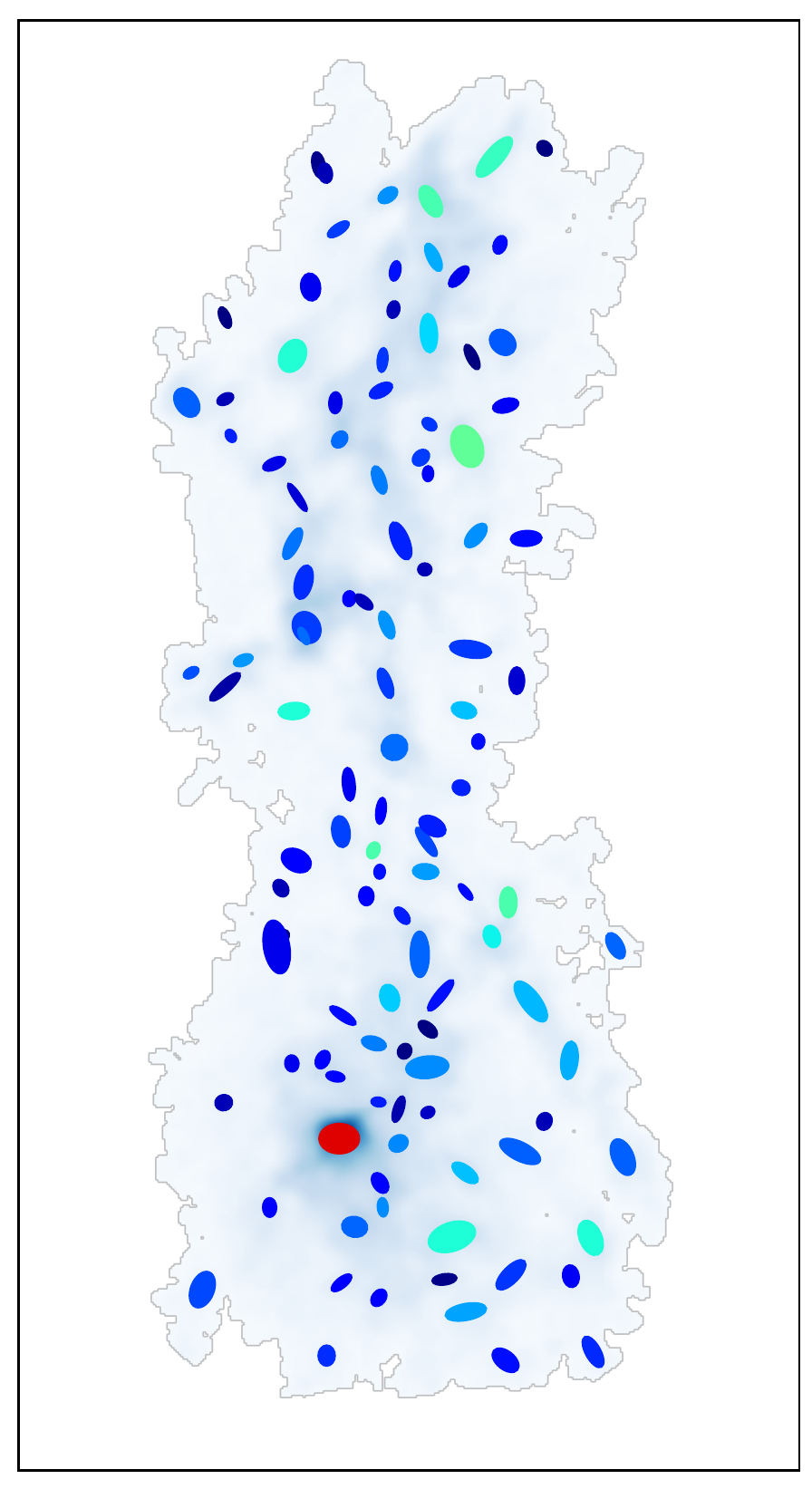}}
  \hfill
  \raisebox{-0.5\height}{\includegraphics[width=4.7in]{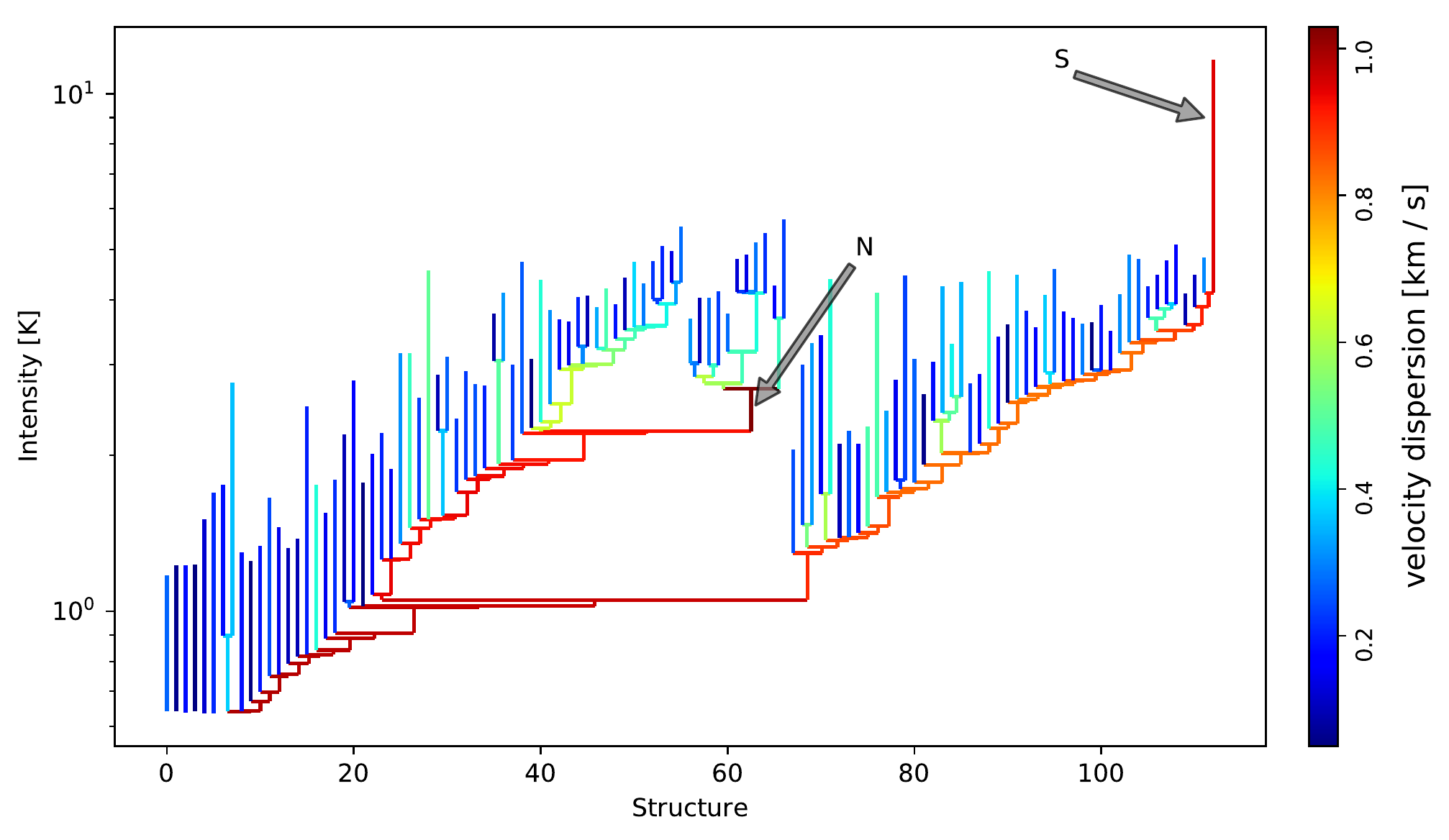}}
\end{minipage}
\begin{minipage}{\textwidth}
  \centering
  \raisebox{-0.5\height}{\includegraphics[width=2.2in, viewport=3 13 476 483,clip=true]{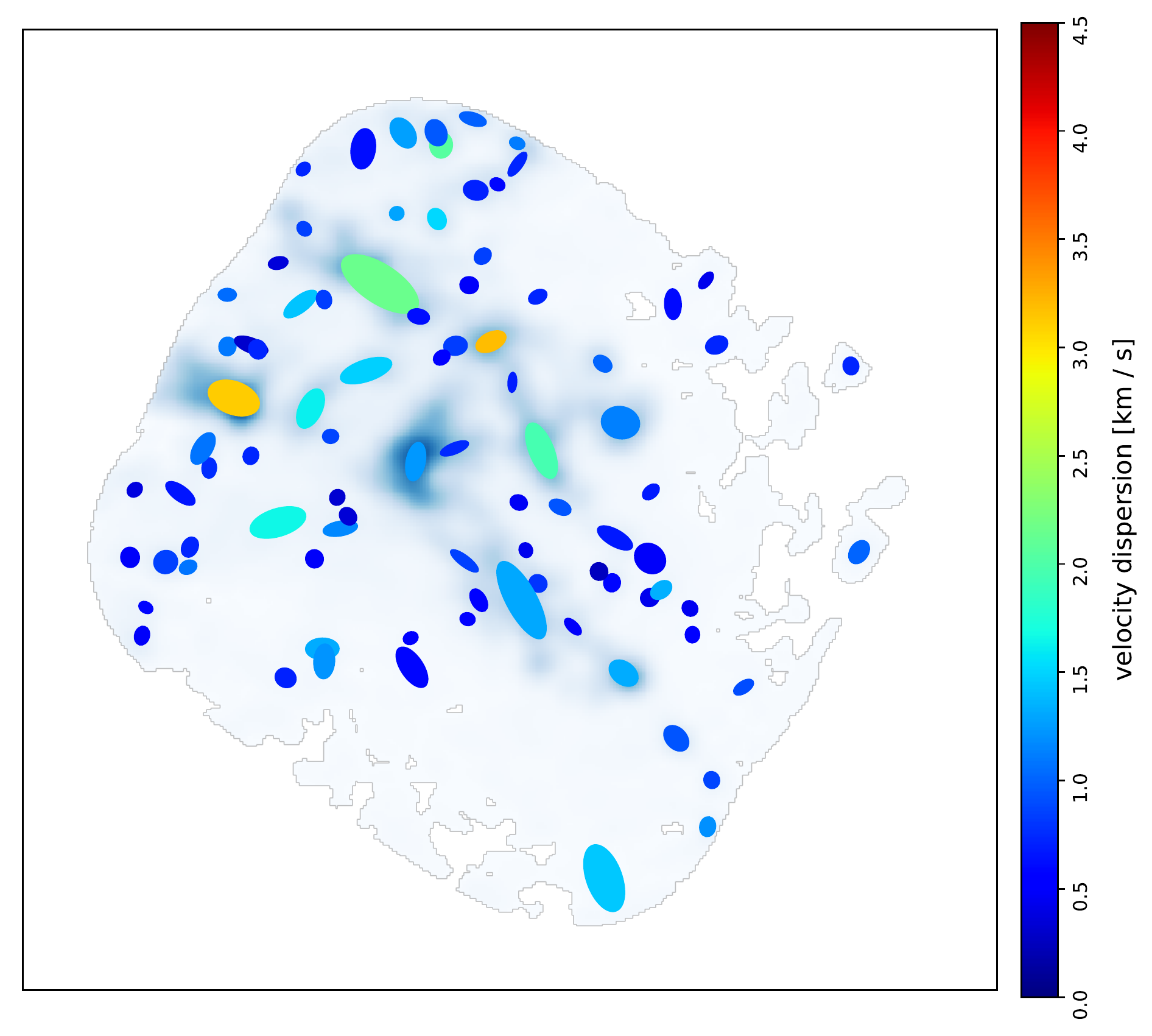}}
  \hfill
  \raisebox{-0.5\height}{\includegraphics[width=4.7in]{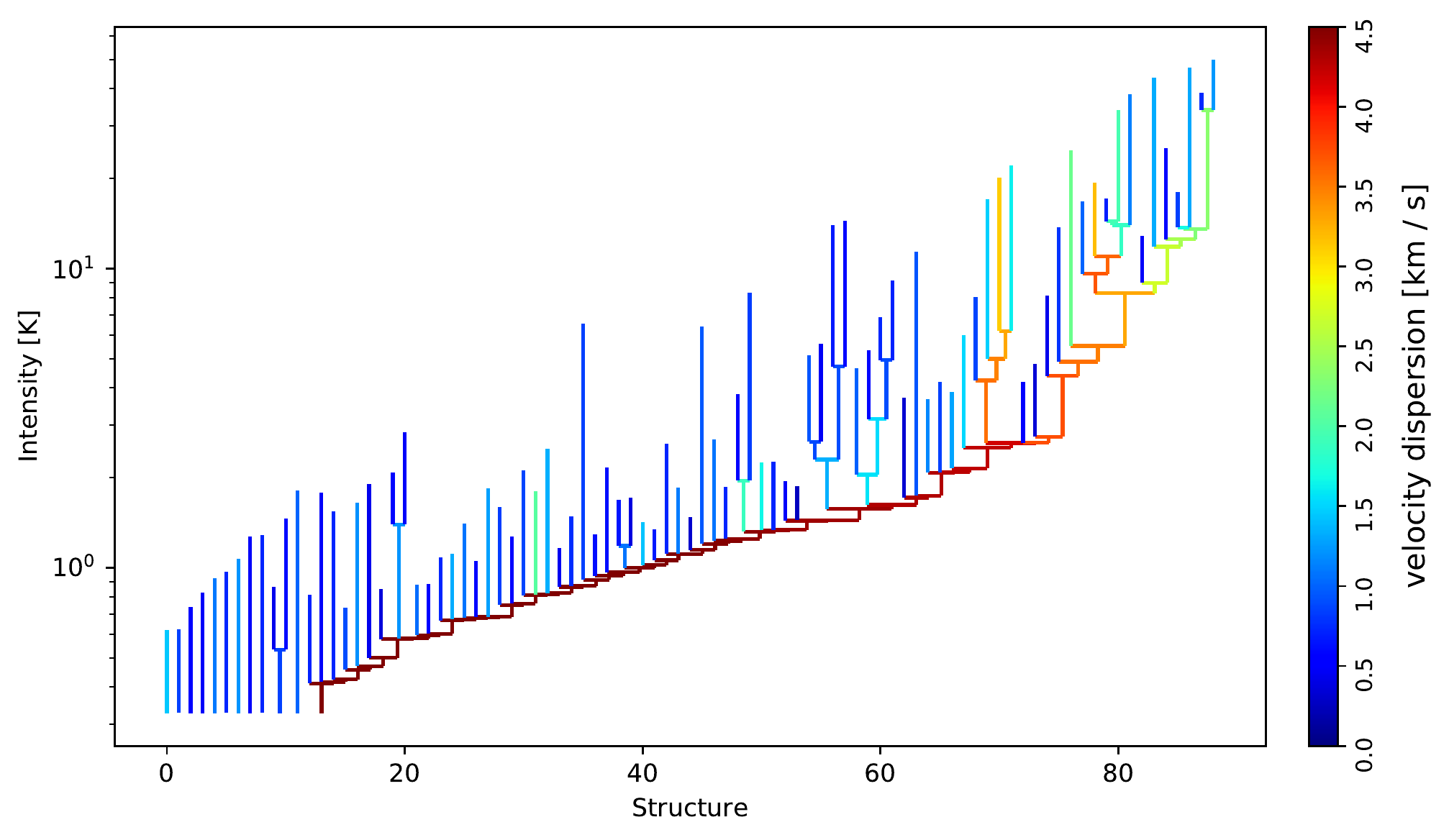}}
\end{minipage}
\caption{\small{$^{12}$CO structures color-coded by velocity dispersion $\sigma_v$.  A schematic of the full dendrogram is shown in the right panels, while the locations and fitted shapes of the dendrogram leaves are shown in the left panels.  The top panels correspond to the PCC, where a single leaf at the location of the bright southern IR source exhibits an exceptionally high $\sigma_v$.  The bottom panels correspond to the 30 Dor cloud.  Arrows in the upper dendrogram indicate the positions for which spectra are displayed in Fig.~\ref{fig:spectra}.
}}
\label{fig:dendro}
\end{figure*}

To identify structures within the ALMA spectral line cubes, we used the Python package {\it astrodendro}, which decomposes emission into a hierarchy of structures \citep{Rosolowsky:08,Shetty:12,Colombo:15}.  Parameters were chosen so that the algorithm identified local maxima in the cube above the 3$\sigma_{\rm rms}$ level that were also at least 2.5$\sigma_{\rm rms}$ above the merge level with adjacent structures.  Each local maximum was required to span at least two synthesized beams in area.  Isosurfaces surrounding the local maxima were categorized as trunks, branches, or leaves according to whether they were the largest contiguous structures (trunks), were intermediate in scale (branches), or had no resolved substructure (leaves).  The resulting dendrograms for CO in PCC and the 30 Dor cloud are shown in the right panels of Figure~\ref{fig:dendro}.

\begin{figure}[tbh]
\begin{center}
\includegraphics[width=0.48\textwidth]{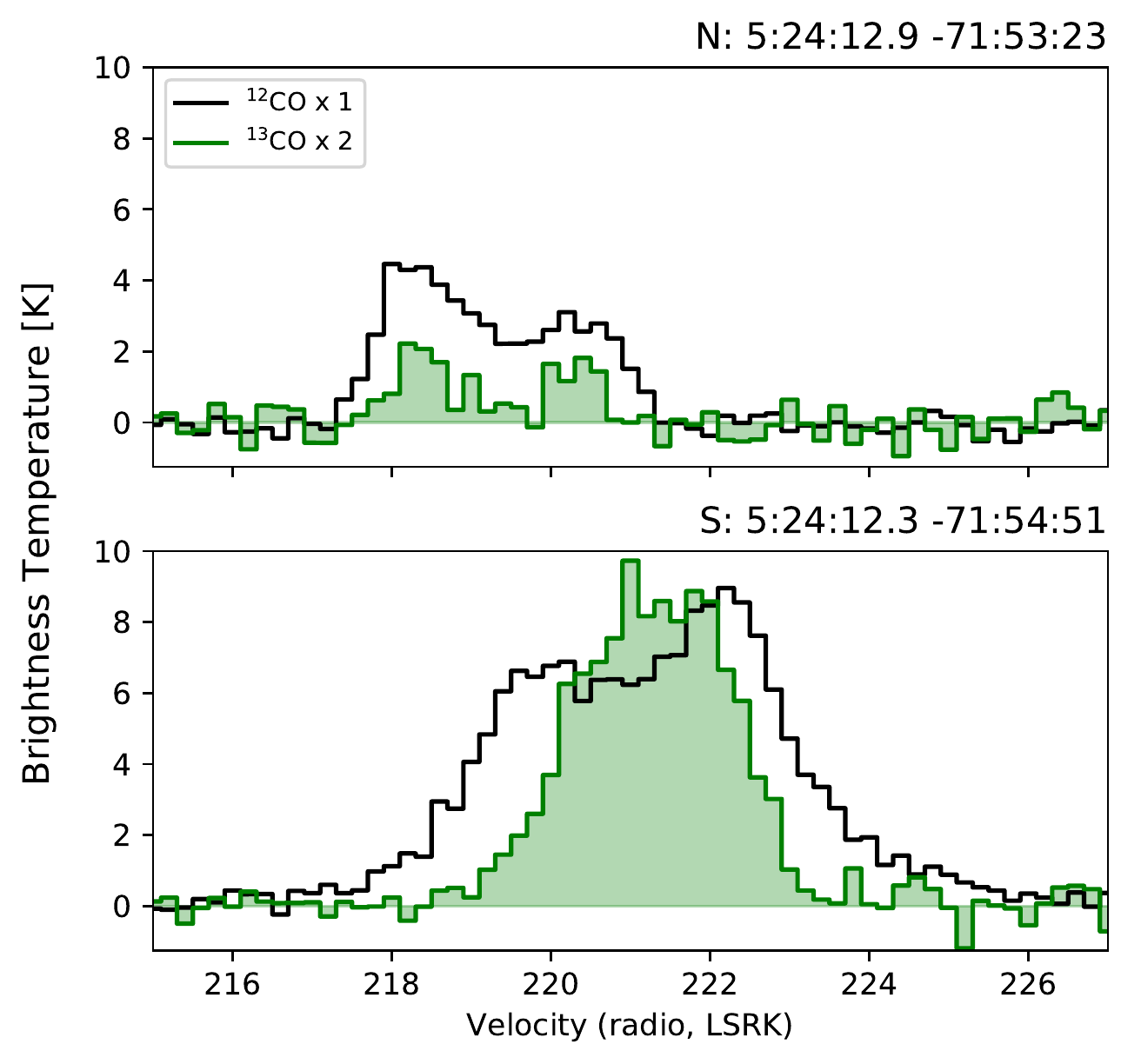}
\end{center}
\caption{CO and $^{13}$CO spectra toward two locations of the PCC indicated by arrows in Fig.~\ref{fig:mom0}. In the northern location the large line width can be attributed to two velocity components whereas the $^{13}$CO profile in the southern location suggests a single, high dispersion component. \label{fig:spectra}}
\end{figure}

The basic properties of the identified structures are also determined by {\it astrodendro}, including their spatial and velocity centroids ($\bar{x}, \bar{y}, \bar{v}$), the integrated flux $S$, the rms line width $\sigma_v$ (defined as the intensity weighted second moment of the structure along the velocity axis), the position angle of the major axis (as determined by principal component analysis) $\phi$, and the rms sizes along the major and minor axes, $\sigma_{\rm maj}$ and $\sigma_{\rm min}$. 
All properties are determined using the ``bijection'' approach discussed by \citet{Rosolowsky:08}, which associates all emission bounded by an isosurface with the identified structure.
From these basic properties we have calculated additional properties, including the effective rms spatial size, $\sigma_r = \sqrt{\sigma_{\rm maj}\sigma_{\rm min}}$, the spherical radius $R = 1.91 \sigma_r$, following \citet{Solomon:87}, the luminosity $L=Sd^2$, adopting $d=48$ kpc \citep{Freedman:10}, the virial mass $M_{\rm vir}=5\sigma_v^2R/G$, derived from solving the equilibrium condition
\begin{equation}\label{eq:vireq}
2{\cal T} + {\cal W} = 2\left(\frac{3}{2}M\sigma_v^2\right) - \frac{3}{5}\frac{GM^2}{R} = 0\,,
\end{equation}
and the luminosity-based mass (for $^{12}$CO)
\[\frac{M_{\rm lum}}{M_\odot} = 4.3X_2\, \frac{L_{\rm CO}}{\rm K\;km\;s^{-1}\;pc^2}\,,\]
where $X_2$=1 for a standard (Galactic) CO to H$_2$ conversion factor \citep{Bolatto:13a}.  In this paper we have adopted $X_2$=4 (scaling the luminosity-based masses upward by a factor of 4) to account for two effects:
\begin{enumerate}
\item The CO(2--1) line is weaker (in $T_b$ units) than the CO(1--0) line by a factor of $\sim$0.6 (\S\ref{sec:ratios}), and our luminosities are for the CO(2--1) line.
\item The CO(1--0) to H$_2$ conversion factor in the LMC is $\sim$2.4 times the factor for the Milky Way, based on the virial analysis of the MAGMA GMC catalog by \citet{Hughes:10}.
\end{enumerate}
In the left panels of Figure~\ref{fig:dendro}, we have used ellipses to denote the centroid positions, position angles, and rms sizes of the dendrogram leaves (the larger branch and trunk structures have been omitted for clarity).  The ellipses have been colored to reflect the rms line width, as discussed further in \S\ref{sec:rdv}.

Following \citet{Rosolowsky:06}, we estimate uncertainties in the derived properties using a bootstrap approach.  These do not take into account the uncertainties in defining the structures, and should be interpreted as the uncertainty in measuring the structure properties assuming the boundaries of the structure in position and velocity space are fixed.  For each structure we generate 100 realizations of the data by random sampling with replacement and use the median absolute deviation to estimate an rms fractional uncertainty for each property.  We scale these uncertainties by the square root of the number of pixels per beam area to account for correlations between pixels.

We do not attempt to correct the sizes and line widths for resolution effects.  As noted by \citet{Rosolowsky:06}, a simple deconvolution approach would underestimate the sizes of structures defined above a given intensity threshold.  An alternative approach would be to extrapolate the sizes and line widths to a zero-intensity contour before deconvolution.  This would complicate the interpretation of nested structures in the dendrogram, however, so we have chosen instead to simply indicate with gray shading the regions where resolution effects become important in our correlation plots.  We note that one consequence of adopting the bijection approach is that the rms line width tends to be underestimated for structures defined by high isocontour levels (i.e., leaves) because the width of the spectral profile is truncated by the isosurface boundary \citep{Rosolowsky:05}.

\begin{figure*}[bth]
\includegraphics[height=2.3in, viewport=3 5 253 314,clip=true]{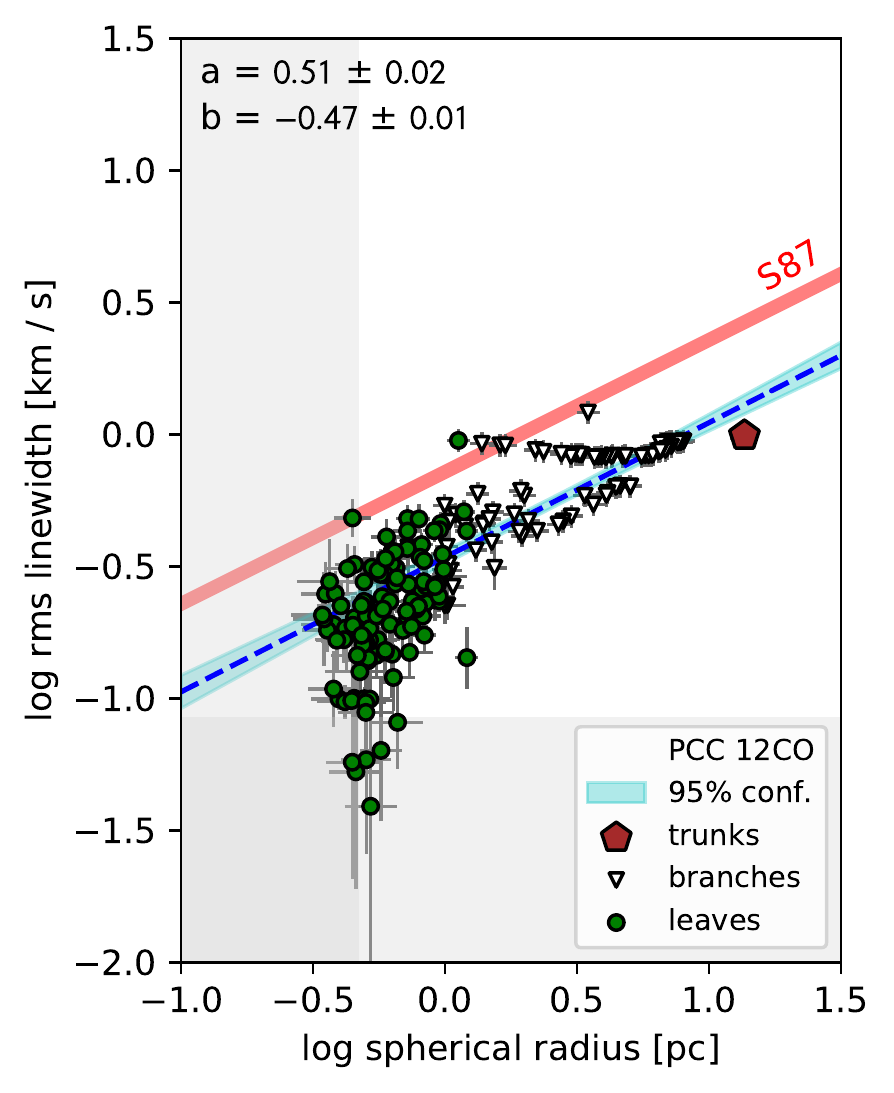}\hfill
\includegraphics[height=2.3in, viewport=19 5 253 314,clip=true]{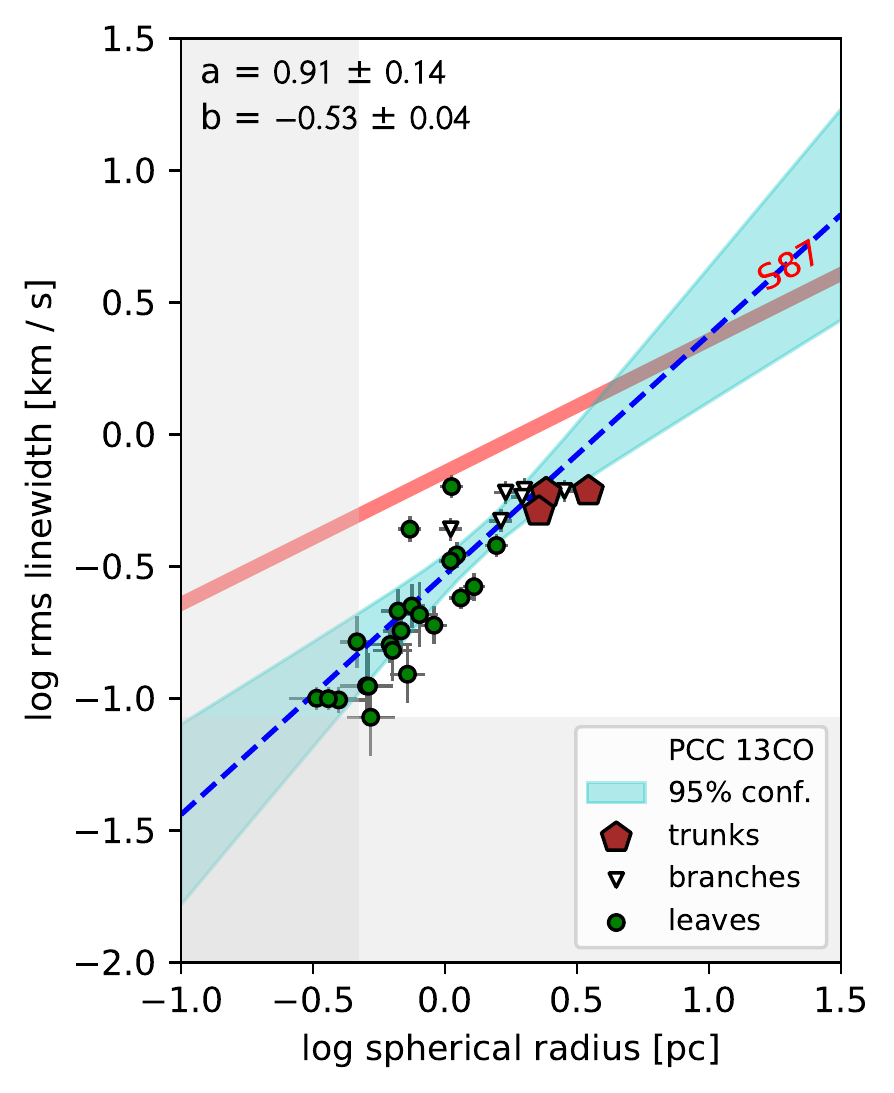}\hfill
\includegraphics[height=2.3in, viewport=19 5 253 314,clip=true]{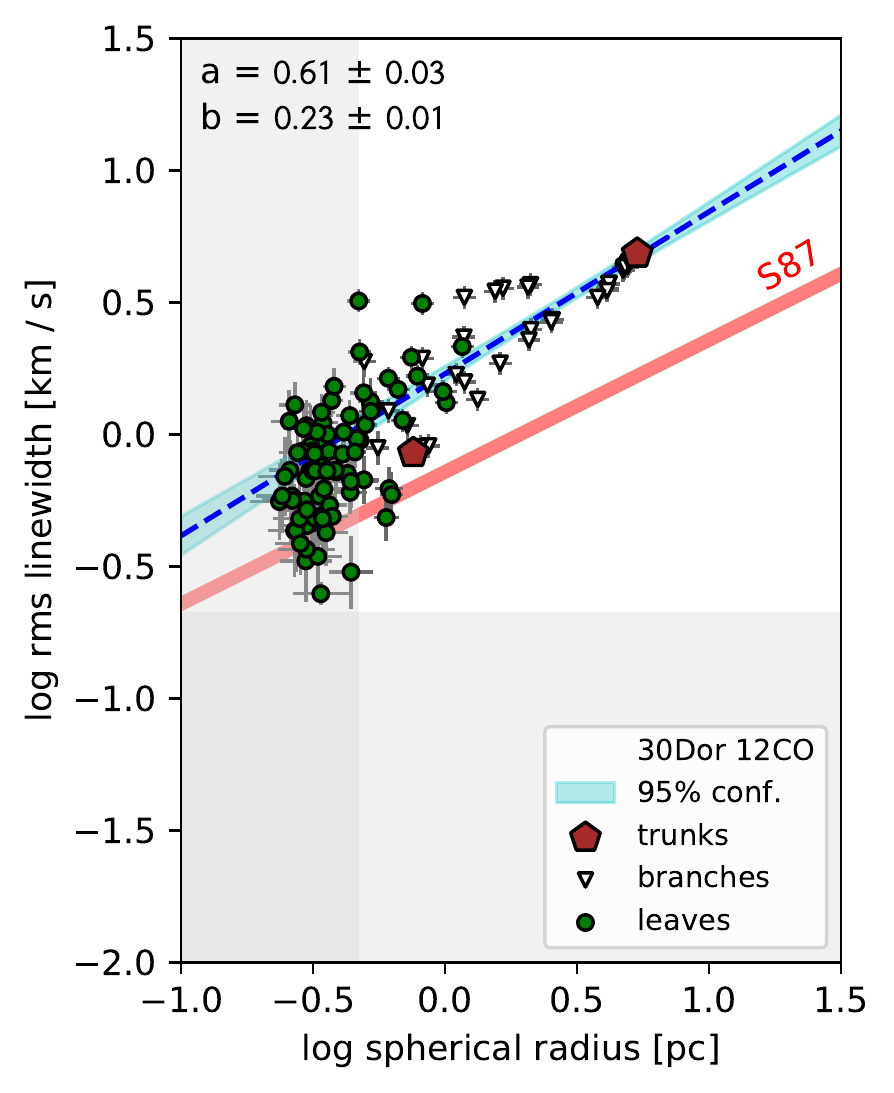}\hfill
\includegraphics[height=2.3in, viewport=19 5 253 314,clip=true]{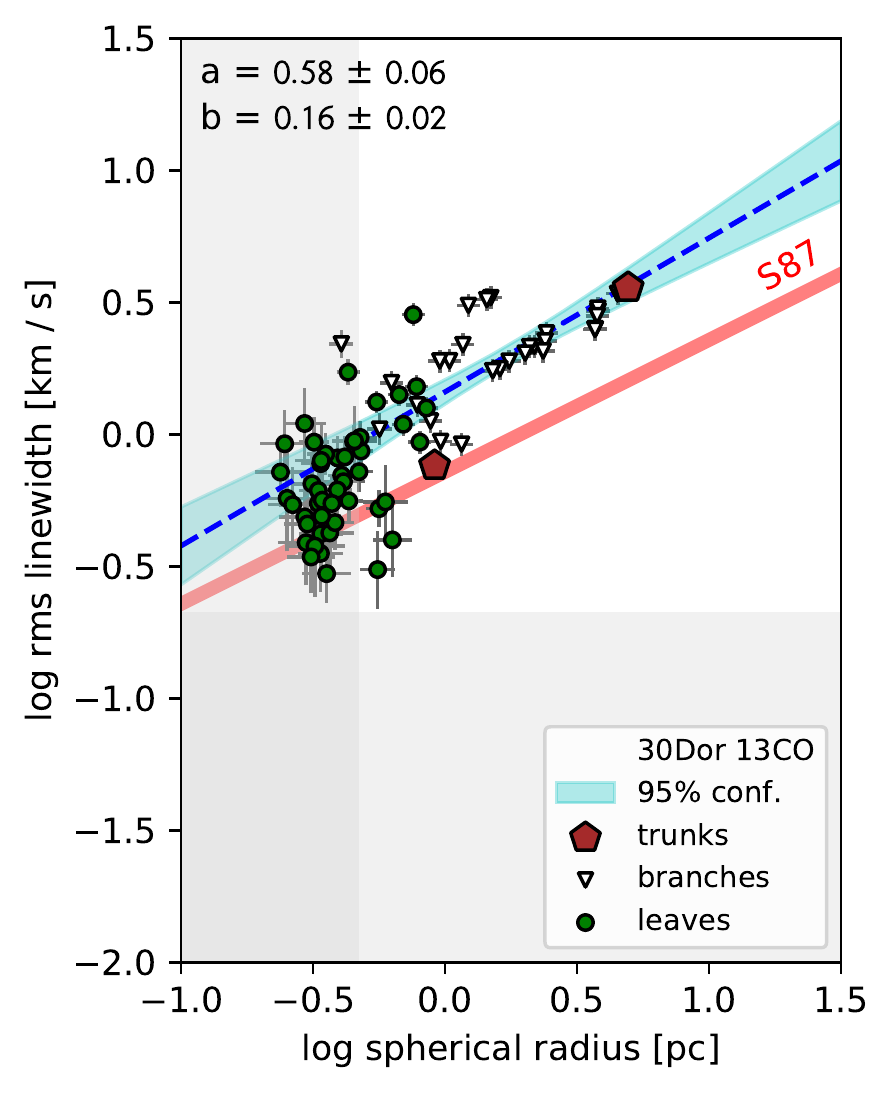}\\
\caption{\small{{\it From left to right}: Size-linewidth relation for $^{12}$CO structures in the PCC, $^{13}$CO structures in the PCC, $^{12}$CO structures in 30 Dor, and $^{13}$CO structures in 30 Dor.  Points are labeled as trunks, branches, or leaves based on their position in the dendrogram.  The Galactic relation from \citet{Solomon:87} is shown as a red line.  Linear fits (blue dashed lines with shaded confidence intervals) are made to all structures except those that fall in the gray shaded regions, which indicate where observational resolution strongly affects the inferred sizes and line widths. 
}}
\label{fig:rdv}
\end{figure*}

In most of the figures presented below we have separately identified leaves, branches, and trunks with different symbols.  Note that the leaves are distinct from each other, as are the trunks, but that all (non-isolated) leaves belong to a trunk.  In contrast, the branches are not independent but lie at intermediate levels of the hierarchy.  

A concern that has been raised about decomposition in position-position-velocity (PPV) cubes is the superposition or blending of physically unrelated structures that occur at the same radial velocity \citep[e.g.,][]{Ostriker:01}.  Assuming that superposition of unrelated structures will sometimes occur at different radial velocities, we can look for sudden changes in line width at different levels of the hierarchy to diagnose such blending.  We find a clear example of such a location as indicated by the arrow labeled `N' in Figure~\ref{fig:mom0}(b), with the corresponding spectra shown in Figure~\ref{fig:spectra} ({\it top}).  Here the similarity of the CO and $^{13}$CO profiles is consistent with superposition, unlike the southern peak labeled `S' [spectra in Fig.~\ref{fig:spectra}, {\it bottom}] where the double-peaked CO profile but single-peaked $^{13}$CO profile suggests the CO profile is self-absorbed.  Such superposition appears to be rare, however, with most of the leaves appearing well-separated spatially.

\subsection{Correlations between cloud properties}\label{sec:corrs}

In order to characterize the relations between structure properties, we fit for linear relationships between the logarithms of properties, i.e.\ power laws.  We employ the {\sc kmpfit} module of the Python package {\tt Kapteyn} \citep{KapteynPackage} which provides non-linear least-squares fitting of user-specified functions.  In order to take into account the estimated errors in both variables, we first obtain an initial estimate of the slope and intercept using an unweighted least-squares fit (code in {\tt scipy.stats.linregress}) and then optimize these parameters in {\sc kmpfit} after weighting each sample by the inverse of its effective variance.  The effective variance method assumes that an error $\delta x_i$ in $x_i$ changes the value of $y_i$ by an amount $f^\prime(x_i)\delta x_i$, which can be added in quadrature to the error $\delta y_i$ in $y_i$.  The {\sc kmpfit} code also provides a standard error (derived from scaling the errors from the covariance matrix to bring the reduced $\chi^2$ to 1) and allows plotting of a 95\% confidence interval.  We exclude from fitting any invalid values for which the logarithm is undefined, and any points which lie below the instrumental resolution, as indicated by the gray shading in the correlation plots.  The resolution in $\sigma_v$ is defined as the channel width divided by $\sqrt{8\ln 2} \approx 2.35$, the resolution in $R$ is defined as the FWHM of the synthesized beam (in this case 2\farcs5 or 0.58 pc) divided by $\sqrt{8\ln 2}$ and multiplied by 1.91, and the resolution in area is defined as the equivalent area of the synthesized beam.

The results of the power-law fits have been summarized in Table~\ref{tab:fitpar}.

\subsubsection{Size-linewidth relations}\label{sec:rdv}

Commonly known as Larson's 1st law, the $R$-$\sigma_v$ relation relates the radius in parsecs to the velocity dispersion in \kms.  \citet{Solomon:87} derived the relation
\[\sigma_v \approx 0.72\, R^{0.5}\;{\rm km\;s^{-1}\;,}\]
which is somewhat flatter than the relations we derive for PCC and 30 Dor (Figure~\ref{fig:rdv}), although still formally within the 3$\sigma$ errors of our fits.  More striking is the variation in the normalization factor: a 1 pc structure in the PCC has a typical line width of 0.3 \kms, whereas a similarly sized structure in 30 Dor has a line width of 1.5 \kms.  While opacity broadening can contribute significantly to the observed line width of $^{12}$CO \citep{Hacar:16b}, the clear separation between PCC and 30 Dor is evident in both CO and $^{13}$CO structures, suggesting that optical depth effects do not strongly affect the normalization.  (On the other hand, opacity effects may flatten the slope of the relation, as evidenced by the steeper slope seen for \ttco\ in the PCC.)  The \citet{Solomon:87} relation has a normalization intermediate between these two regions, consistent with PCC and 30 Dor representing the two extremes of star formation activity in the LMC.

\begin{figure}[bh]
\includegraphics[width=0.45\textwidth]{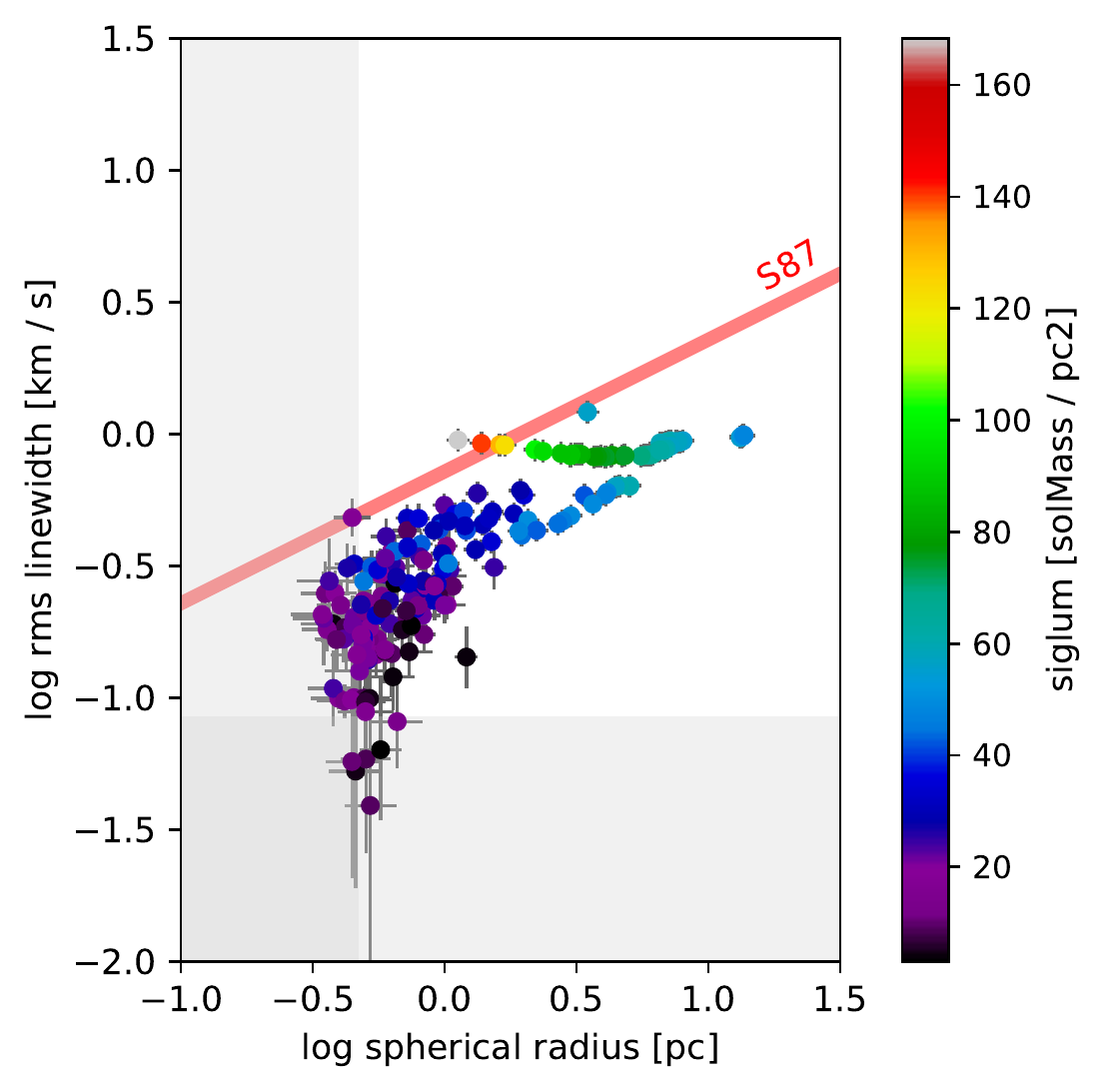}
\caption{\small{Size-linewidth relation for $^{12}$CO structures in the PCC, color coded by CO surface brightness (expressed as an equivalent H$_2$ surface density). 
}}
\label{fig:rdvcolor}
\end{figure}

The other important insight from Figure~\ref{fig:rdv} is that the spread in $\sigma_v$ is already quite large on the smallest scales of the dendrogram hierarchy, i.e.\ the leaves indicated as green filled circles.  Specifically, $\sigma_v$ for leaves ranges up to 0.95 \kms\ for the PCC (3.20 \kms\ for 30 Dor), whereas the highest measured $\sigma_v$ is 1.21 \kms\ (4.84 \kms\ for 30 Dor).  The high-dispersion leaves show up distinctly in the color coding of Figure~\ref{fig:dendro} ({\it left panels}): in the case of the PCC, the CO peak labeled as `S' in Figure~\ref{fig:mom0}(b) dominates the line widths of all leaves in both CO and $^{13}$CO.  This CO peak is coincident with an infrared (IR) point source that is discussed further in \S\ref{sec:core}.  The sequence of high-dispersion branches that connects this leaf to its underlying trunk is clearly visible in the left panel of Figure~\ref{fig:rdv} and as the red lines at the lower right of the dendrogram tree.  The nearly constant $\sigma_v$ in these branches (with a slight decline) reflects the strong influence of the high dispersion in that single bright leaf.  {\it The $R$-$\sigma_v$ relation is thus driven in part by the tendency for large structures to inherit the largest line widths of their constituents.}  This is accentuated by the tendency for high-dispersion structures to exhibit higher CO surface brightness (Figure~\ref{fig:rdvcolor}), and thus dominate the properties of the enclosing structures in the hierarchy.  Since small structures with narrow line widths exert little influence on the line widths of the larger structures they are embedded in, the upper envelope of the size-linewidth plot tends to be much flatter than the lower envelope.

At the same time, we still find that the largest values of $\sigma_v$ result from merging small-linewidth structures exhibiting different line-of-sight velocities; the location labeled as `N' in Figure~\ref{fig:mom0}(b) is a good example of this phenomenon, which is seen in the dendrogram (Figure~\ref{fig:dendro}) as leaves with low dispersion (colored blue) merging into branches with high dispersion (colored red).  Furthermore, since the vast majority of leaves have low dispersion, there is still a general trend of increasing dispersion with increasing size scale.  These features are characteristic of a turbulent spectrum spanning a range of scales with increasing kinetic energy on larger scales.  We discuss the implications of clouds exhibiting a handful of high-dispersion leaves along with a general increase in dispersion with size scale further in \S\ref{sec:scimes}.

\subsubsection{Mass-radius relations}\label{sec:mrad}

\begin{figure*}[tbh]
\includegraphics[width=0.25\textwidth]{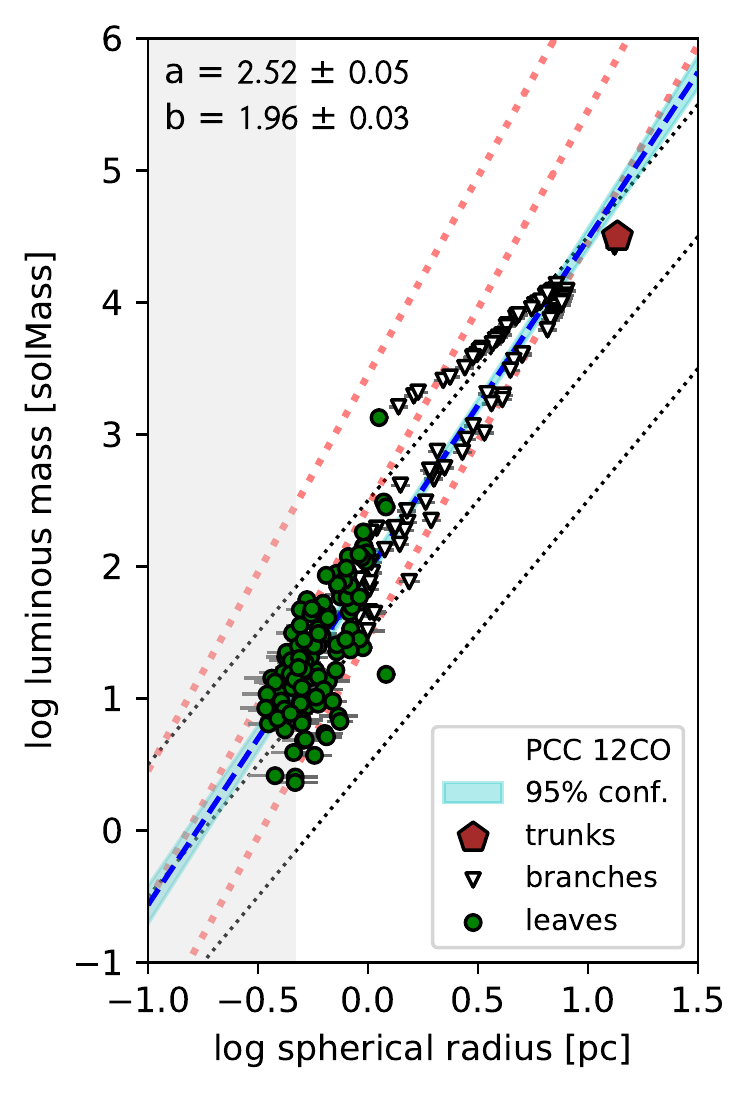}\hfill
\includegraphics[width=0.25\textwidth]{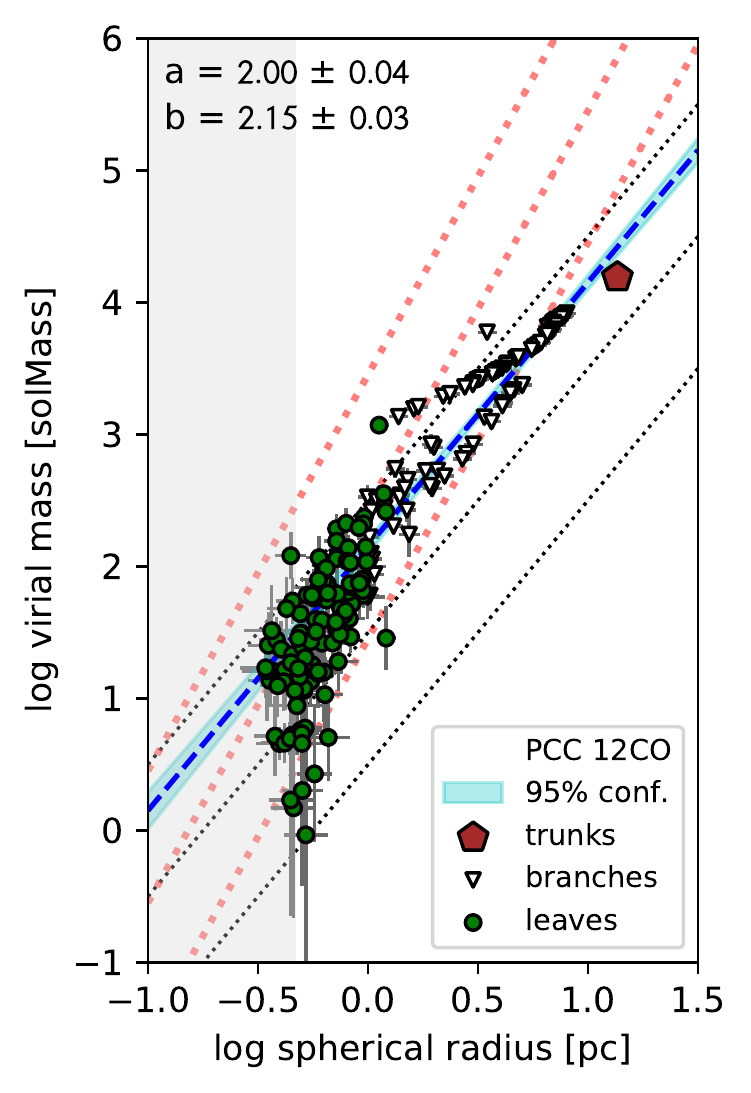}\hfill
\includegraphics[width=0.25\textwidth]{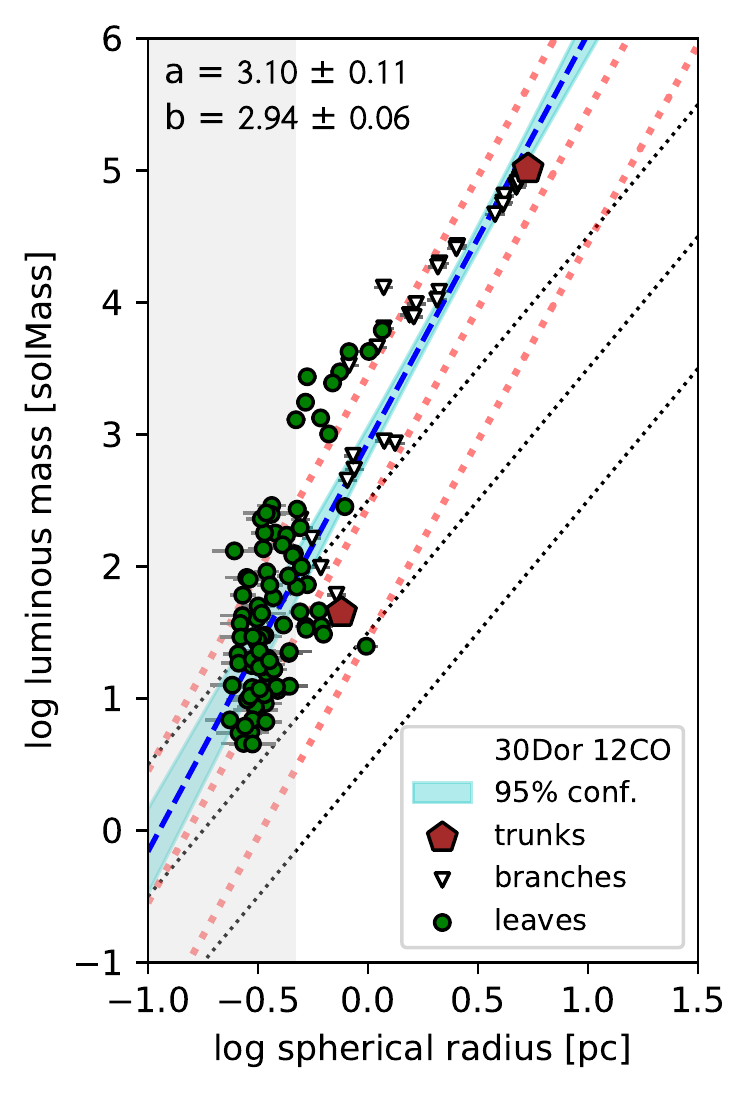}\hfill
\includegraphics[width=0.25\textwidth]{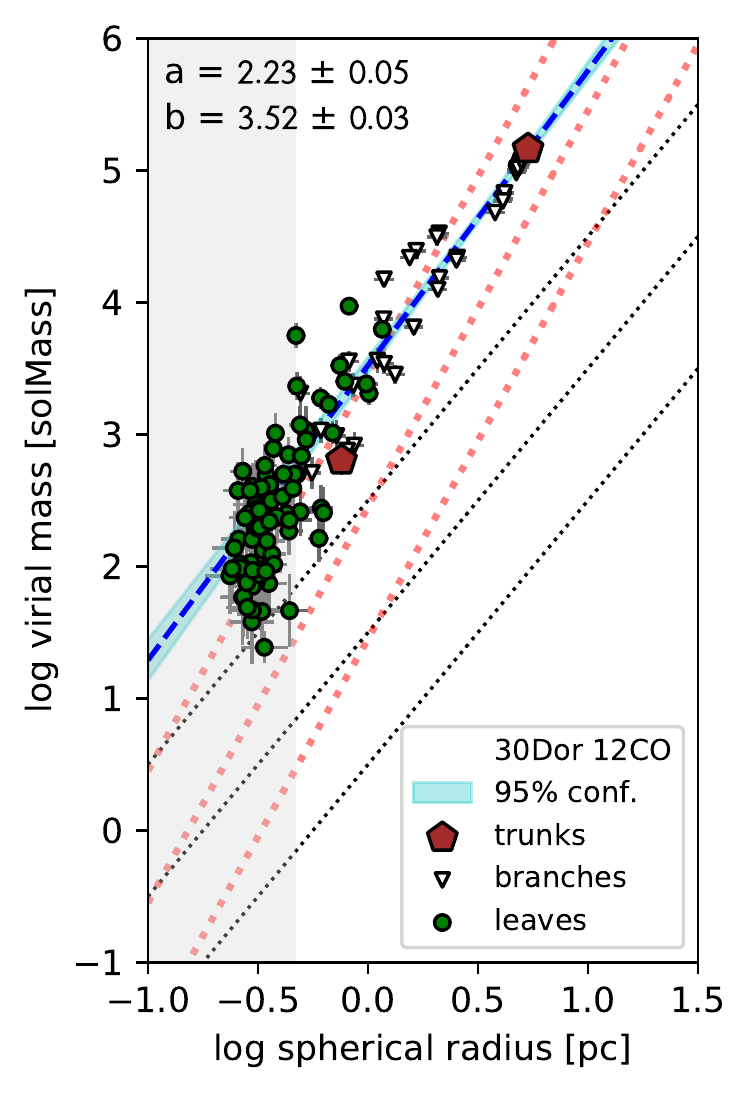}\\
\caption{\small{Mass-radius relations for the PCC (first two panels, using luminous and virial mass indicators respectively) and 30 Doradus (last two panels), shown for the CO-identified structures.  The dotted black lines indicate surface densities of 1, 10, and 100 \Msol\ pc$^{-2}$, while dotted pink lines indicate average volume densities of $10^2$, $10^3$, and $10^4$ molecules cm$^{-3}$.  The dashed blue line represents the best-fit power-law relationship with slope ($a$) and intercept ($b$) given at the upper left.
}}
\label{fig:areaflux}
\end{figure*}

Figure~\ref{fig:areaflux} shows the correlation of two CO mass estimators (integrated CO flux assuming a constant $X_{\rm CO}$ factor and virial mass) with the equivalent spherical radius $R$.  Diagonal black dotted lines show average surface densities of 1, 10, and 100 \Msol\ pc$^{-2}$, while red dotted lines show average volume densities of $n({\rm H}_2) = 10^2$, $10^3$, and $10^4$ cm$^{-3}$.  For comparison, the 1$\sigma$ CO surface brightness sensitivity of the ALMA maps (at 0.58 pc resolution) are 0.18 and 0.4 K \kms\ for PCC and 30 Dor respectively, corresponding to a 3$\sigma$ surface density of 9.2 and 21 \Msol\ pc$^{-2}$.  Some of the smallest structures are close to this limit, but larger structures are well above it.  Note that some measured surface densities may fall below the nominal limit if they are derived from averaging over several beams, or over a smaller range in velocity than the moment mask.

We find that for a given mass estimator, clumps in 30 Dor show about an order of magnitude higher CO surface brightness than those in the PCC, reflecting a combination of higher brightness temperature and larger line width.  For the PCC, the typical surface densities (based on virial mass) are comparable to or slightly less than the canonical value of 206 \Msol\ pc$^{-2}$ derived by \citet{Solomon:87} for Galactic clouds, also using virial masses, whereas the surface densities in 30 Dor are much higher, close to $10^3$ \Msol\ pc$^{-2}$ in many cases.  The volume densities show similar differences, being typically $10^2$--$10^3$ cm$^{-3}$ for the PCC and $\sim 10^4$ cm$^{-3}$ for 30 Dor.

We also find that larger structures tend to show a narrower range of surface density, as expected from ensemble averaging, with the highest surface densities found at intermediate size scales.  The latter property may be related to the fact that structures are essentially identified as isocontour levels in the cube.  Small isocontours can enclose both high and low brightness regions, while intermediate-sized isocontours enclose relatively bright regions that span multiple contour levels, and the largest isocontours enclose the bulk of the emission in the cube, averaging over moderate and high-brightness regions.

We note that the limited field-of-view of the ALMA mosaics confines our observations to regions of bright CO emission---the relatively high surface densities on large scales are thus partly a result of targeting the observations toward the CO peaks of low-resolution single-dish maps.
When it comes to other properties, however, our comparisons with the wide-field MAGMA data indicate overall consistency between the ALMA-observed regions and the larger CO cloud.  For instance, while the ALMA 30 Dor mosaic only covers a portion of the cloud identified as 30Dor-10 by \citet{Johansson:98}, the line width of the entire cloud (identified as A331 in \citealt{Wong:11}) is 4.73 \kms, similar to the value of 4.84 \kms\ we derive on the largest scale of the ALMA map.  The mass of the entire cloud, with our adopted $X_{\rm CO}$, is about $1.75 \times 10^5$ \Msol, compared with $1.0 \times 10^5$ \Msol\ for the region covered by the ALMA map.

\subsubsection{Luminosity-mass ($L$-$M$) relations} 

\begin{figure*}[tbh]
\begin{center}
\includegraphics[width=0.45\textwidth]{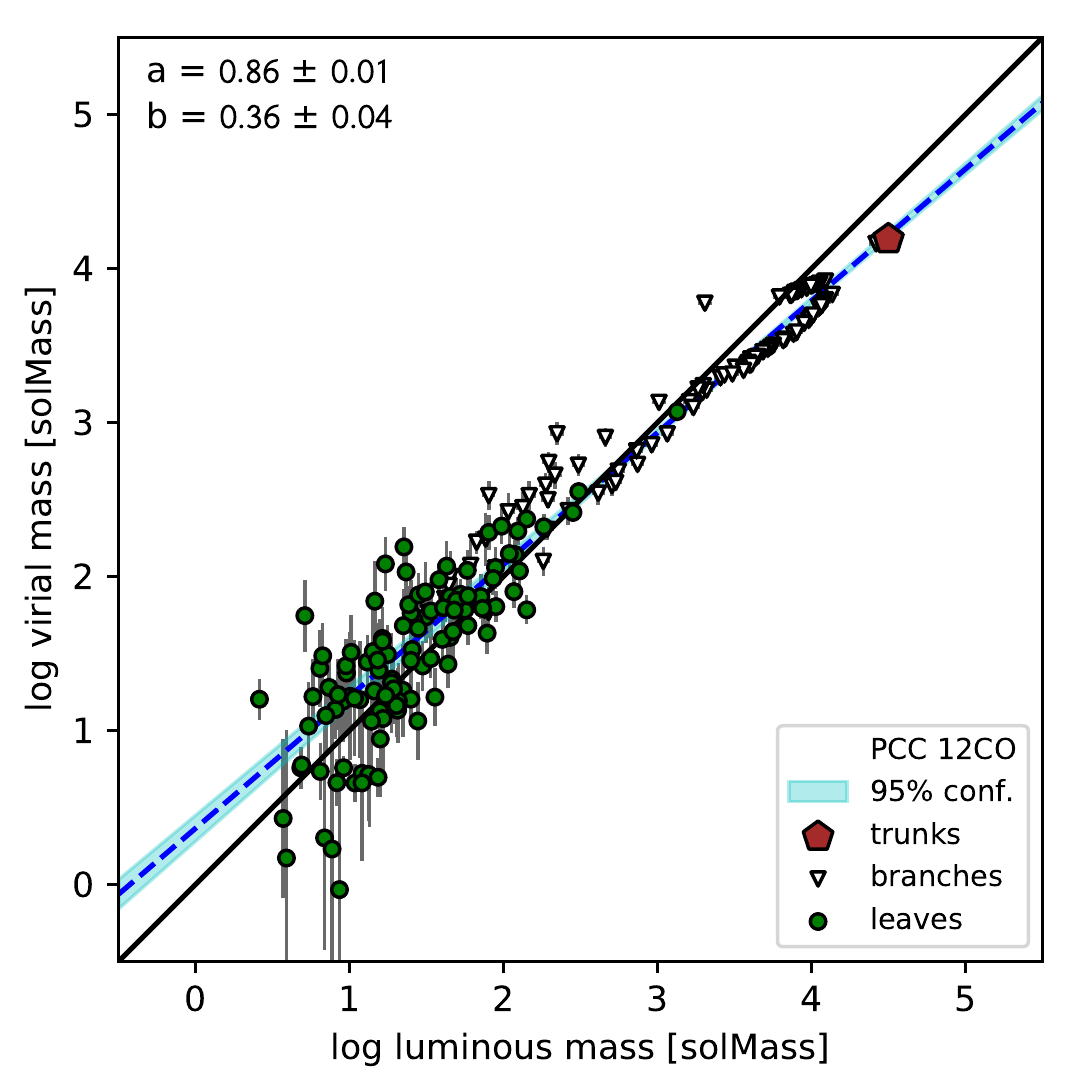}\hspace*{1cm}
\includegraphics[width=0.45\textwidth]{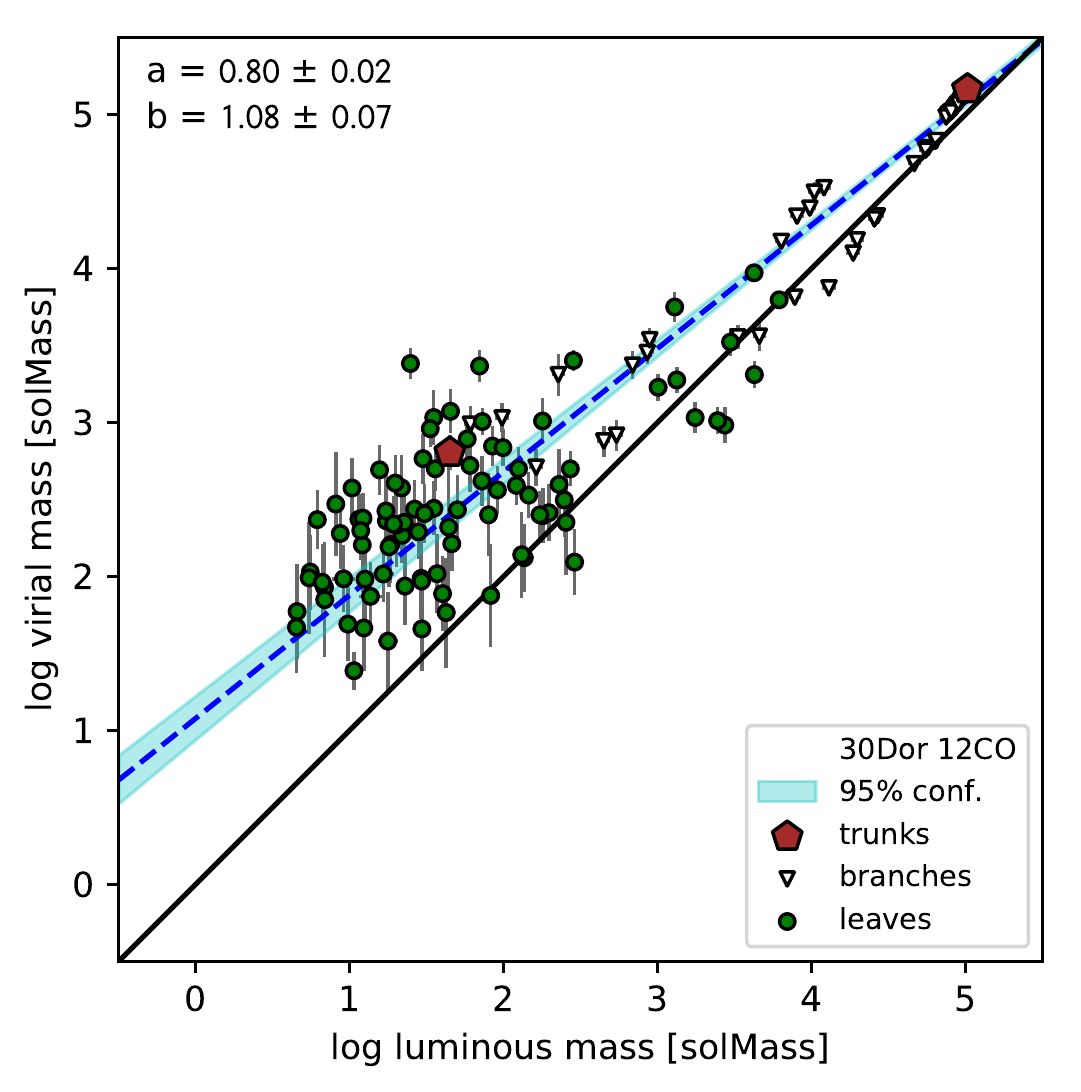}\\
\end{center}
\caption{\small{Relation between virial mass and luminosity-based mass (assuming a constant $\alpha_{\rm CO}$ factor as described in the text) and virial mass for \twco-identified structures in the PCC ({\it left}) and 30 Doradus ({\it right}). 
}}
\label{fig:mlumvir}
\end{figure*}

Figure~\ref{fig:mlumvir} shows the correlation between two \twco-based mass estimators: the ``luminous mass,'' based on the integrated CO flux assuming a constant $X_{\rm CO}$ factor, and the virial mass.  
With the $X_{\rm CO}$ factor we have adopted in this paper (\S\ref{sec:dendro}), the two masses appear to be roughly comparable in the PCC, whereas the virial mass tends to be larger in 30 Dor, particularly for the smallest structures.  Since $M_{\rm vir} \propto R\sigma_v^2$, this partially reflects the larger line widths observed in 30 Dor.
In both clouds, the slope of the correlation is significantly flatter than unity, indicating that more massive structures are more strongly bound (i.e.\ possess a smaller virial parameter $\alpha_{\rm vir} \equiv M_{\rm vir}/M_{\rm lum}$).  As noted by \citet{Shetty:10}, the slope of the $M_{\rm lum}$--$M_{\rm vir}$ relation is related to the slopes of the $R$--$\sigma_v$ and mass-radius correlations; if $M_{\rm lum} \propto R^\alpha$ and $\sigma_v \propto R^\beta$ then $M_{\rm vir} \propto M_{\rm lum}^{2\beta+1/\alpha}$.  The sub-linear slope ($2\beta+1 < \alpha$) arises because the $R$--$\sigma_v$ relation is relatively flat (as noted in \S\ref{sec:rdv}, larger structures tend to inherit the largest line widths from their constituents) whereas the $R$--$M_{\rm lum}$ relation is relatively steep (\S\ref{sec:mrad}; larger structures encompass much more mass than their constituents).  
We caution, however, that opacity effects in \twco\ may contribute to the flatness of the $R$--$\sigma_v$ relation, as discussed in \S\ref{sec:rdv}.

\subsubsection{$L$-$M$ relations normalized by area} \label{sec:bnd}

\begin{figure*}[tbh]
\begin{center}
\includegraphics[width=0.5\textwidth]{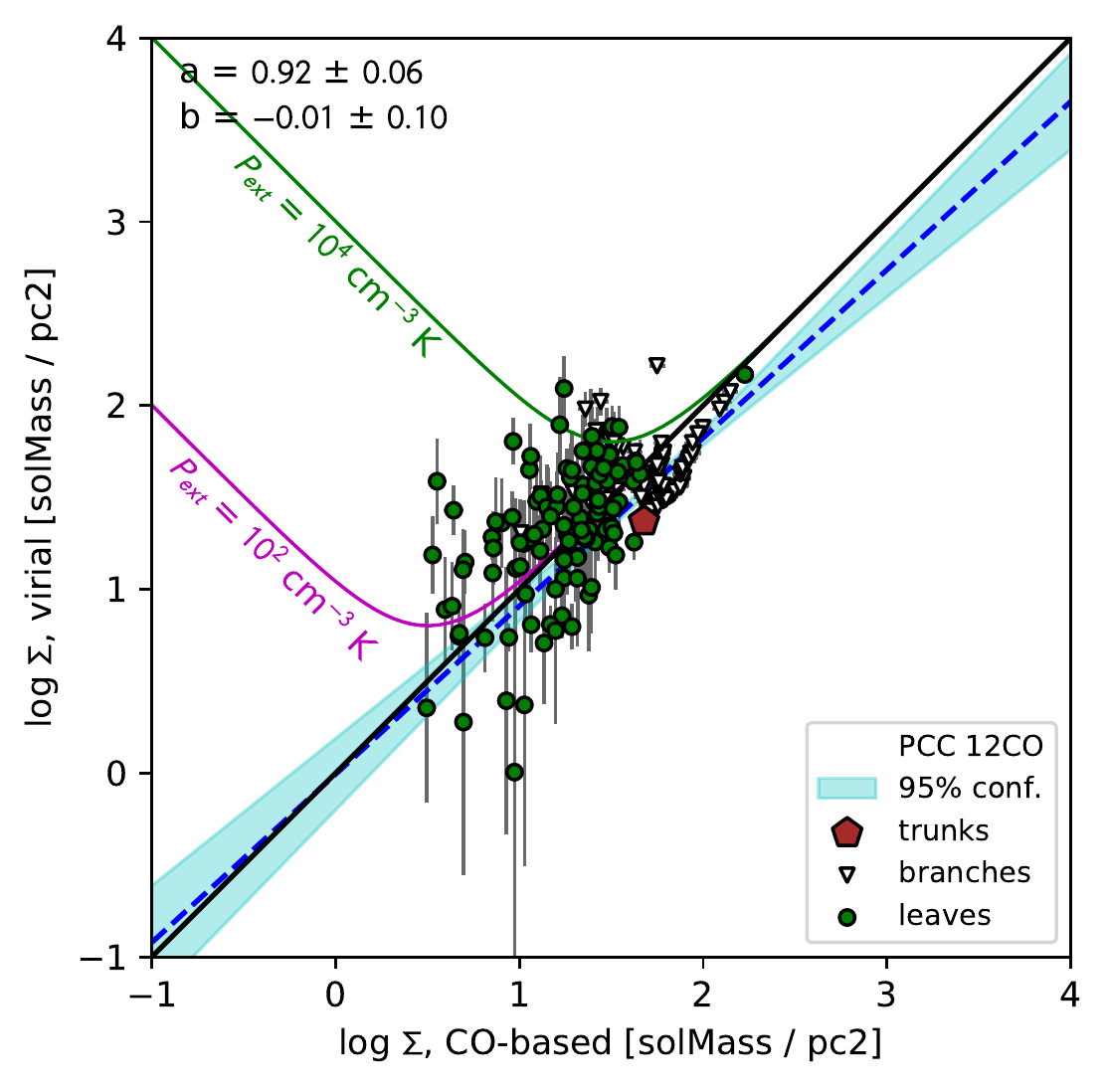}\hfill
\includegraphics[width=0.5\textwidth]{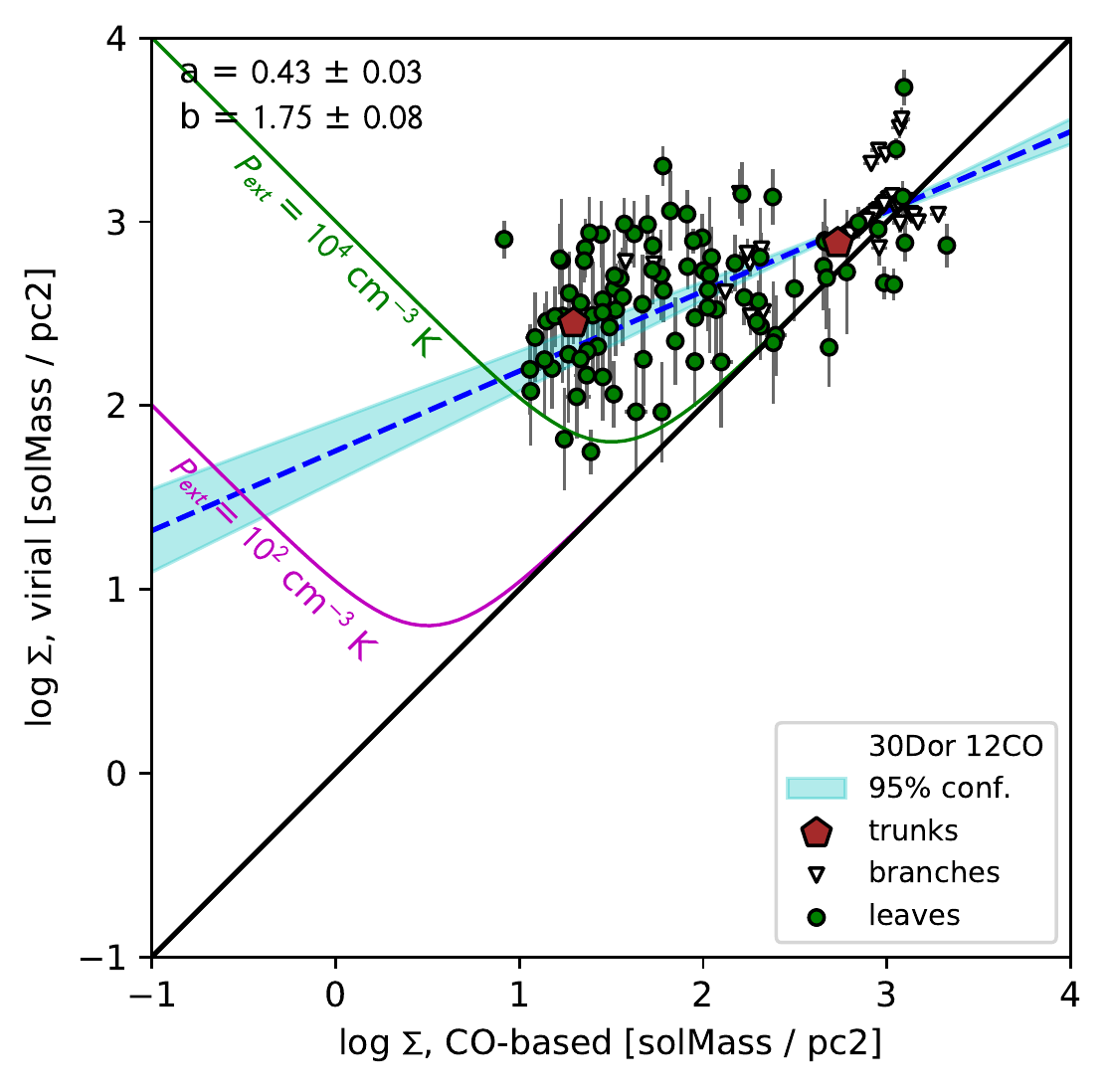}\\
\end{center}
\caption{\small{Relation between virial and CO-based surface densities for \twco-identified structures in the PCC ({\it left}) and 30 Doradus ({\it right}).  The line of slope unity represents simple virial equilibrium, while green and magenta curves represent loci of pressure-bounded equilibria.
}}
\label{fig:bnd}
\end{figure*}

In Figure~\ref{fig:bnd} the virial and luminous masses have been normalized by the projected area $A$ of each structure to yield mass surface densities, which are then compared.
Since $M_{\rm vir} \propto R\sigma_v^2$, this is essentially a plot of $\sigma_v^2/R$ against $\Sigma$, which is analogous to the ``boundedness'' plot discussed by \citet{Heyer:09} (see also \citealt{Keto:86}, \citealt{Field:11}, \citealt{Leroy:15}).  This plot removes much of the intrinsic correlation in the $L$-$M$ plot and allows loci of pressure-bounded equilibrium to be distinguished (at least in principle) from virialization.
The equality line represents the simple virial equilibrium (SVE) relation of Equation~(\ref{eq:vireq}), while the two curves represent pressure-bounded equilibria with external pressures of $P_{\rm ext}/k = 10^2$ and $10^4\rm\; cm^{-3}\, K$, derived from setting the left-hand side of Equation~(\ref{eq:vireq}) equal to $4\pi R^3 P_{\rm ext}$:
\begin{equation}
\Sigma_{\rm vir} - \Sigma_{\rm lum} = \frac{20}{3\pi G}\frac{P_{\rm ext}}{\Sigma_{\rm lum}}\;.
\end{equation}
Overall, the observed structures, particularly in the PCC, lie close to SVE, although the higher virial-to-luminous mass ratios observed in 30 Dor are again apparent in this figure.  There is a population of lower surface-density leaves in both clouds that lie well above the SVE line, 
and are thus consistent with pressure confinement, albeit not at a single value of $P_{\rm ext}$.  Especially for 30 Dor, the offset from the SVE line is often much larger than the factor of $\sim$2 (0.3 dex) or $\sqrt{2}$ (0.15 dex) that has been attributed to pressure-bounded virial equilibrium \citep{Field:11} or energy equipartition due to free-fall collapse \citep{Vazquez:07,Ballesteros:11a} respectively, disfavoring those explanations.  It is important to note that in this context ``lower surface density'' is defined in a relative sense for each cloud, since the out-of-equilibrium structures seen in 30 Dor are found at surface densities where most structures appear to be virialized in the PCC.

\begin{deluxetable*}{CCCRCRCCCR}
\tablehead{
\colhead{} & \colhead{} & \multicolumn{2}{c}{PCC \twco} & \multicolumn{2}{c}{PCC \ttco} & \multicolumn{2}{c}{30\,Dor \twco} & \multicolumn{2}{c}{30\,Dor \ttco}\\
\cline{3-4} \cline{5-6} \cline{7-8} \cline{9-10}
\colhead{$X$} & \colhead{$Y$} & \colhead{$a$} & \colhead{$b$} & \colhead{$a$} & \colhead{$b$} & \colhead{$a$} & \colhead{$b$} & \colhead{$a$} & \colhead{$b$}} 
\tablecaption{Power Law Fit Parameters: $\log Y = a \log X + b$\label{tab:fitpar}}
\decimals
\startdata
R & $\sigma_v$ & 0.51 \pm 0.02 & -0.47 \pm 0.01 & 0.91 \pm 0.14 & -0.53 \pm 0.04 & 0.61 \pm 0.03 & 0.23 \pm 0.01 & 0.58 \pm 0.06 & 0.16 \pm 0.02 \\
R & $M_{\rm lum}$ & 2.52 \pm 0.05 & 1.96 \pm 0.03 & \nodata & \nodata & 3.10 \pm 0.11 & 2.94 \pm 0.06 & \nodata & \nodata \\
R & $M_{\rm vir}$ & 2.00 \pm 0.04 & 2.15 \pm 0.03 & 2.89 \pm 0.29 & 1.98 \pm 0.07 & 2.23 \pm 0.05 & 3.52 \pm 0.03 & 2.20 \pm 0.12 & 3.38 \pm 0.05 \\
$M_{\rm lum}$ & $M_{\rm vir}$ & 0.86 \pm 0.01 & 0.36 \pm 0.04 & \nodata & \nodata & 0.80 \pm 0.02 & 1.08 \pm 0.07 & \nodata & \nodata \\
$\Sigma_{\rm lum}$ & $\Sigma_{\rm vir}$ & 0.92 \pm 0.06 & -0.01 \pm 0.10 & \nodata & \nodata & 0.43 \pm 0.03 & 1.75 \pm 0.08 & \nodata & \nodata \\
$M_{\rm lum}$ & $M_{\rm LTE}$ & 0.95 \pm 0.01 & 0.53 \pm 0.03 & \nodata & \nodata & 0.97 \pm 0.01 & 0.44 \pm 0.05 & \nodata & \nodata \\
$M_{\rm vir}$ & $M_{\rm LTE}$ & 1.04 \pm 0.02 & 0.39 \pm 0.08 & 0.94 \pm 0.05 & 0.99 \pm 0.13 & 1.06 \pm 0.02 & -0.14 \pm 0.11 & 1.00 \pm 0.03 & 0.35 \pm 0.14 \\
\enddata
\end{deluxetable*}

\subsection{Comparison with $^{13}$CO-based masses}\label{sec:lte}

In concluding that molecular clouds and their clumps are close to virial equilibrium, there is a risk of circularity given that our luminosity-to-mass conversion ($X_2$=4) was calibrated in part by assuming virial equilibrium.  We use \ttco\ as an alternative mass tracer assuming that \twco\ and \ttco\ are in local thermodynamic equilibrium (LTE) at a common excitation temperature \citep[e.g.,][]{Nishimura:15}.
We assume the $^{12}$CO (2--1) line is optically thick at line center and not subject to beam dilution, so that for a given line of sight the excitation temperature is uniform and given by
\begin{equation}
T_{\rm ex}{\rm [K]} = 11.06 \left[\ln\left(1+\frac{11.06}{T_{\rm 12,pk}+0.187}\right)\right]^{-1}\;,
\end{equation}
where $T_{\rm 12,pk}$ is the peak temperature of the \twco\ line profile.  The $^{13}$CO (2--1) optical depth is then calculated from the brightness temperature $T_{13}$ at each position and velocity in the cube using
\begin{equation}
\tau_{13} = -\ln\left[1-\frac{T_{13}}{10.6}\left\{\frac{1}{e^{10.6/T_{\rm ex}}-1}-\frac{1}{e^{10.6/2.7}-1}\right\}^{-1}\right]\;.
\end{equation}
Since $\tau_{13}$ varies linearly with $T_{13}$ in the optically thin limit, we do not exclude $T_{13}$ values below a 3$\sigma$ sensitivity limit, but instead allow negative values of $\tau_{13}$ due to noise; given the limited range of $T_{\rm ex}$ (as discussed below), these tend to average out when integrated in the cube.

\begin{figure*}[tbh]
\begin{center}
\includegraphics[width=0.45\textwidth]{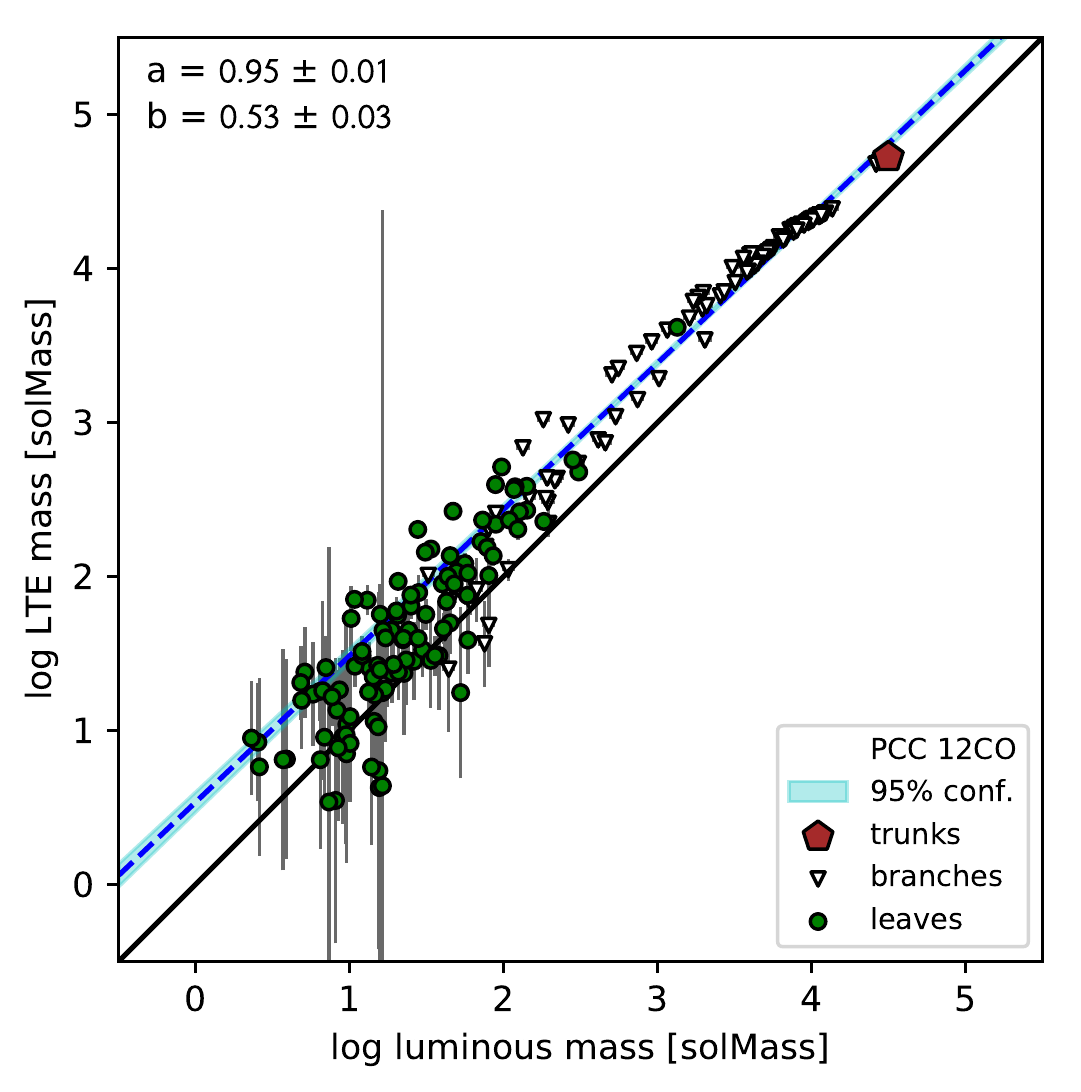}\hspace*{1cm}
\includegraphics[width=0.45\textwidth]{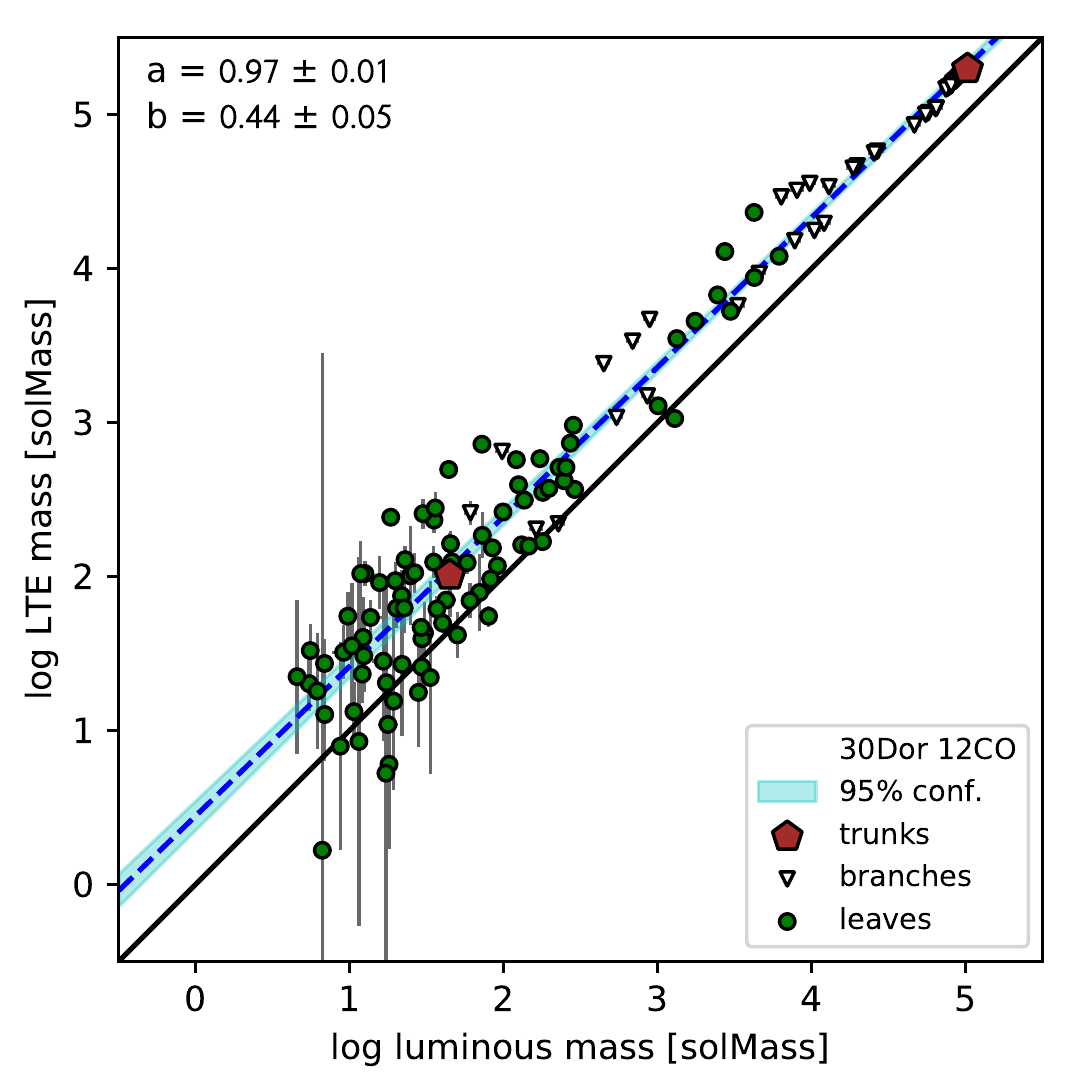}\\
\includegraphics[width=0.45\textwidth]{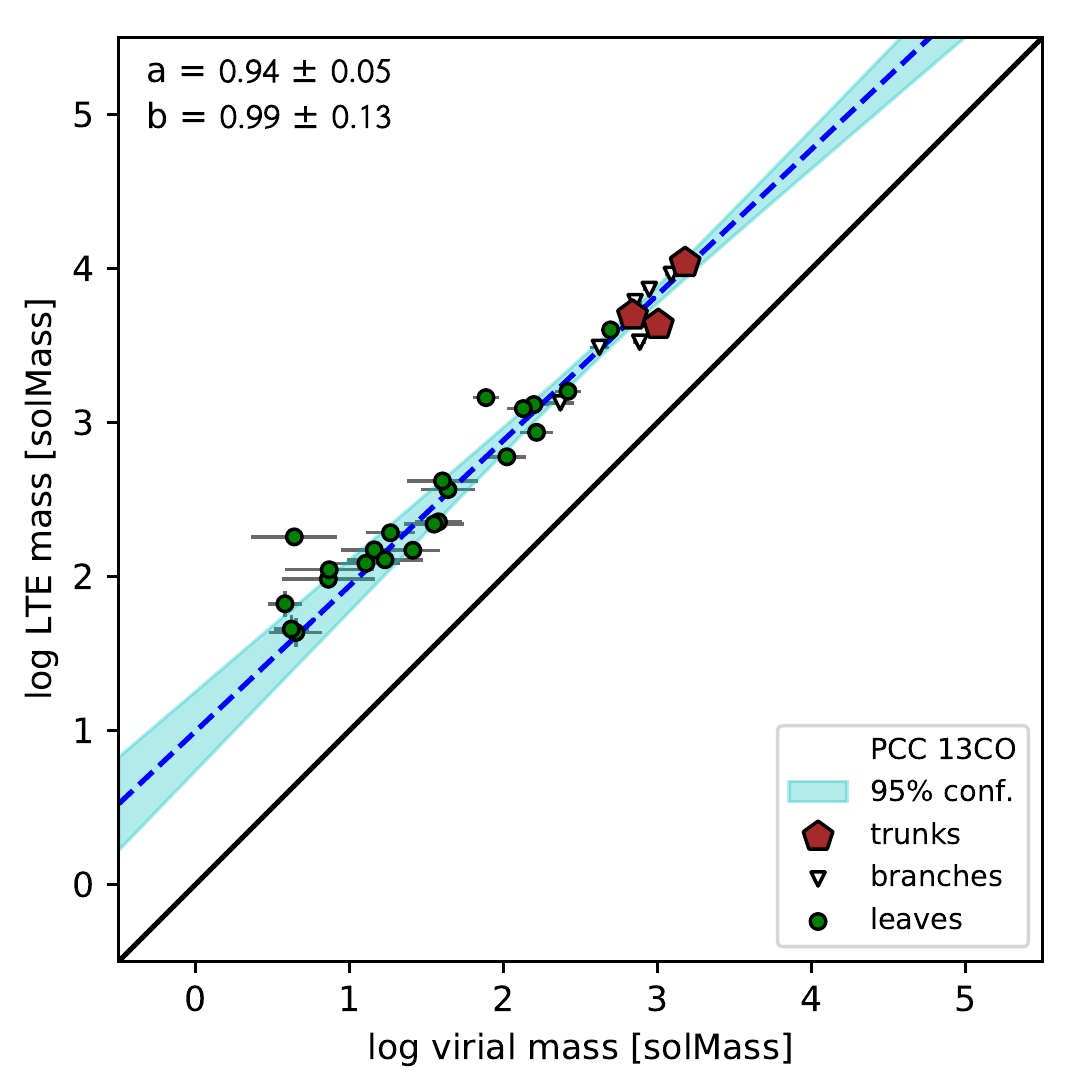}\hspace*{1cm}
\includegraphics[width=0.45\textwidth]{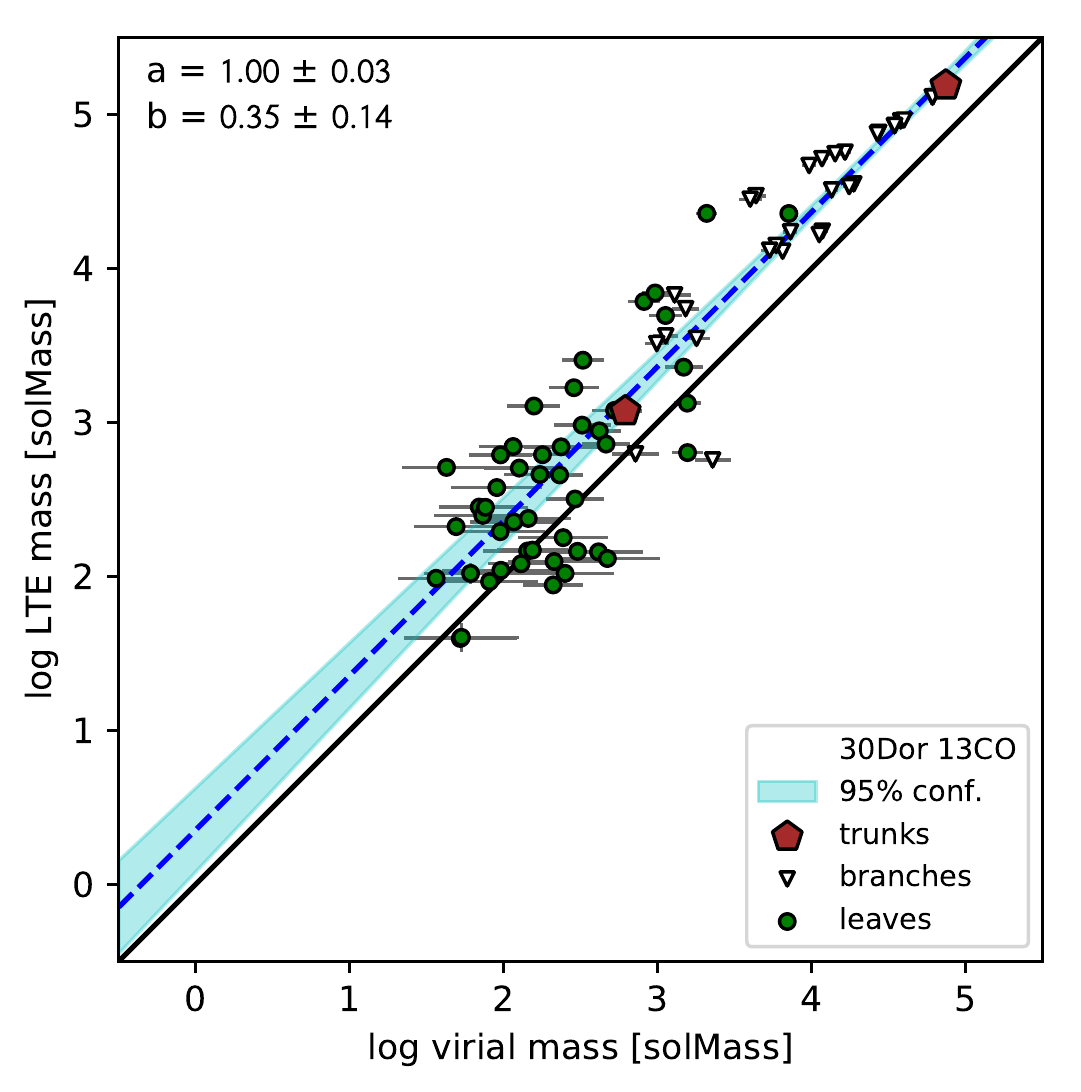}\\
\end{center}
\caption{\small{{\it Top panels}: Relation between luminosity-based mass (assuming a constant $\alpha_{\rm CO}$ factor) and LTE-based mass for \twco-identified structures in the PCC ({\it left}) and 30 Doradus ({\it right}). {\it Bottom panels}: Relation between virial mass and LTE-based mass for \ttco-identified structures in the PCC ({\it left}) and 30 Doradus ({\it right}).
}}
\label{fig:masscorr}
\end{figure*}

For any given $T_{\rm ex}$, there is a maximum allowed value of $T_{13}$ beyond which $\tau_{13}$ becomes undefined.  For $T_{\rm ex}$=8 K we require $T_{13}<3.6$ K.  To reduce the number of such undefined values we impose a minimum ``floor'' of 8 K on $T_{\rm ex}$ under the assumption that lower inferred values reflect beam dilution.  Still, given that locally we observe values of $T_{13}$ as high as $\sim$7 K, there are a handful of 3D pixels for which the $^{13}$CO column density cannot be calculated; these pixels are excluded from the LTE mass estimates, which might therefore be biased small for the most luminous structures.  
Applying a $T_{\rm ex}$ floor significantly reduces the number of undefined values of $\tau_{13}$, although too high a floor value reduces the allowed range of $T_{\rm ex}$.  At 2\farcs5 resolution, we derive values of $T_{\rm ex}$ ranging up to 17 K for the PCC and 57 K for 30 Dor.  The total $^{13}$CO column density in cm$^{-2}$, summed over all rotational levels, is then given by \citep{Garden:91,Bourke:97}:
\begin{equation}
N(^{13}{\rm CO}) = 1.21 \times 10^{14}\frac{(T_{\rm ex}+0.88)e^{5.29/T_{\rm ex}}\int \tau_{13}\,dv}{1-e^{-10.6/T_{\rm ex}}}\;.
\end{equation}

Errors were estimated using a Monte Carlo approach, with many realizations of the input $^{12}$CO and $^{13}$CO cubes generated, consistent with the noise determined from signal-free channels, to determine a resultant uncertainty map for $N$($^{13}$CO).  For each realization the $T_{\rm ex}$ floor was randomly varied between 6 and 10 K.  In practice the uncertainty was calculated channel by channel in order to allow the column density measurement to be restricted to a 3D mask.  We stress that these uncertainties mainly reflect the uncertainties due to map noise. 
The assumption of a single $T_{\rm ex}$ that can be derived from $T_{\rm 12,pk}$, as well as the definition of the 3D masks used for integration, introduce additional systematic uncertainties.  
Table~\ref{tab:regs} shows clearly, for instance, that beam dilution reduces $T_{\rm 12,pk}$ when smoothing to a common resolution of 2\farcs5; our resulting underestimate of $T_{\rm ex}$ would lead us to overestimate $\tau_{13}$ and $N$($^{13}$CO).

\begin{figure*}[tbh]
\includegraphics[width=0.48\textwidth]{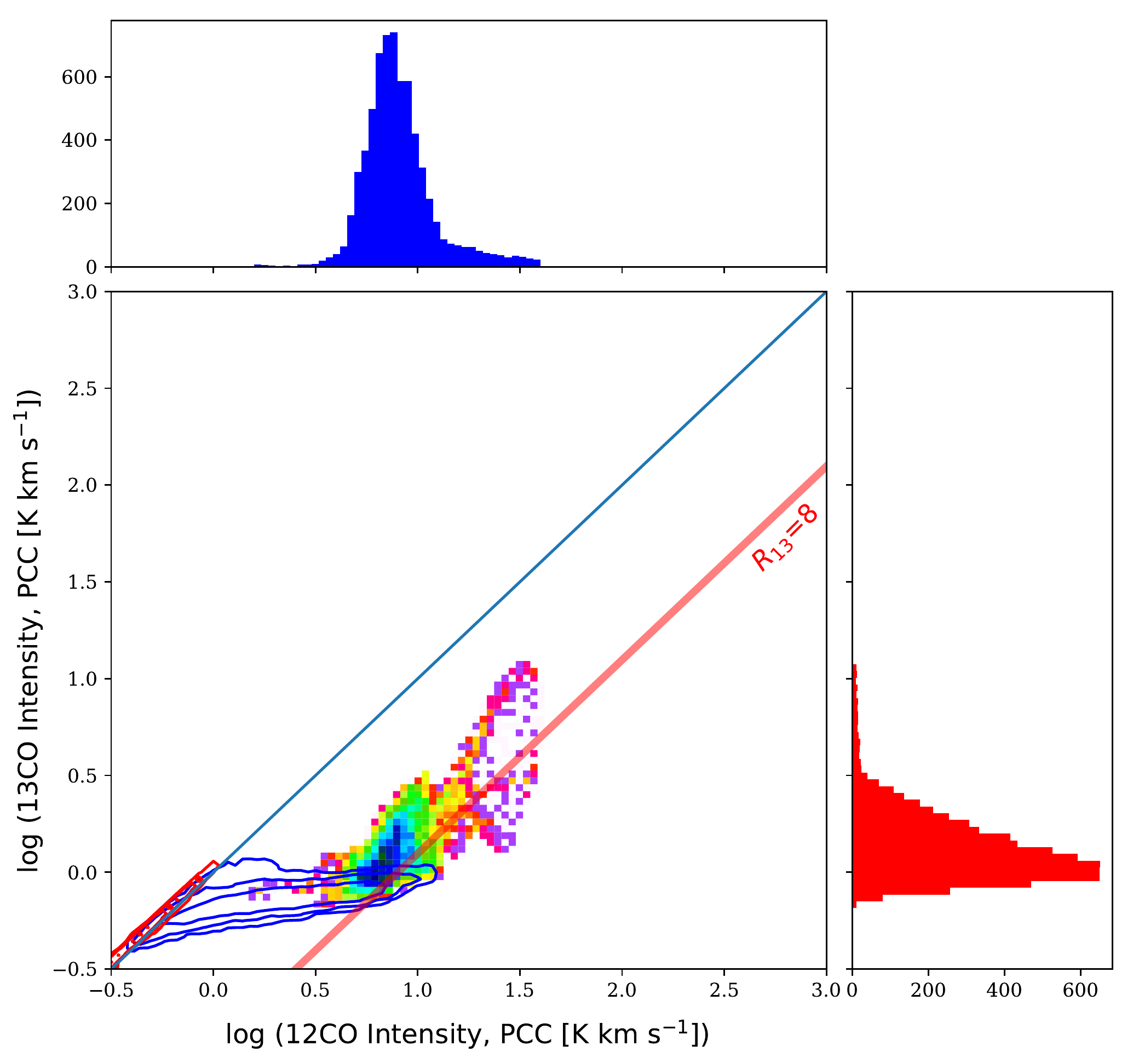}\hfill
\includegraphics[width=0.48\textwidth]{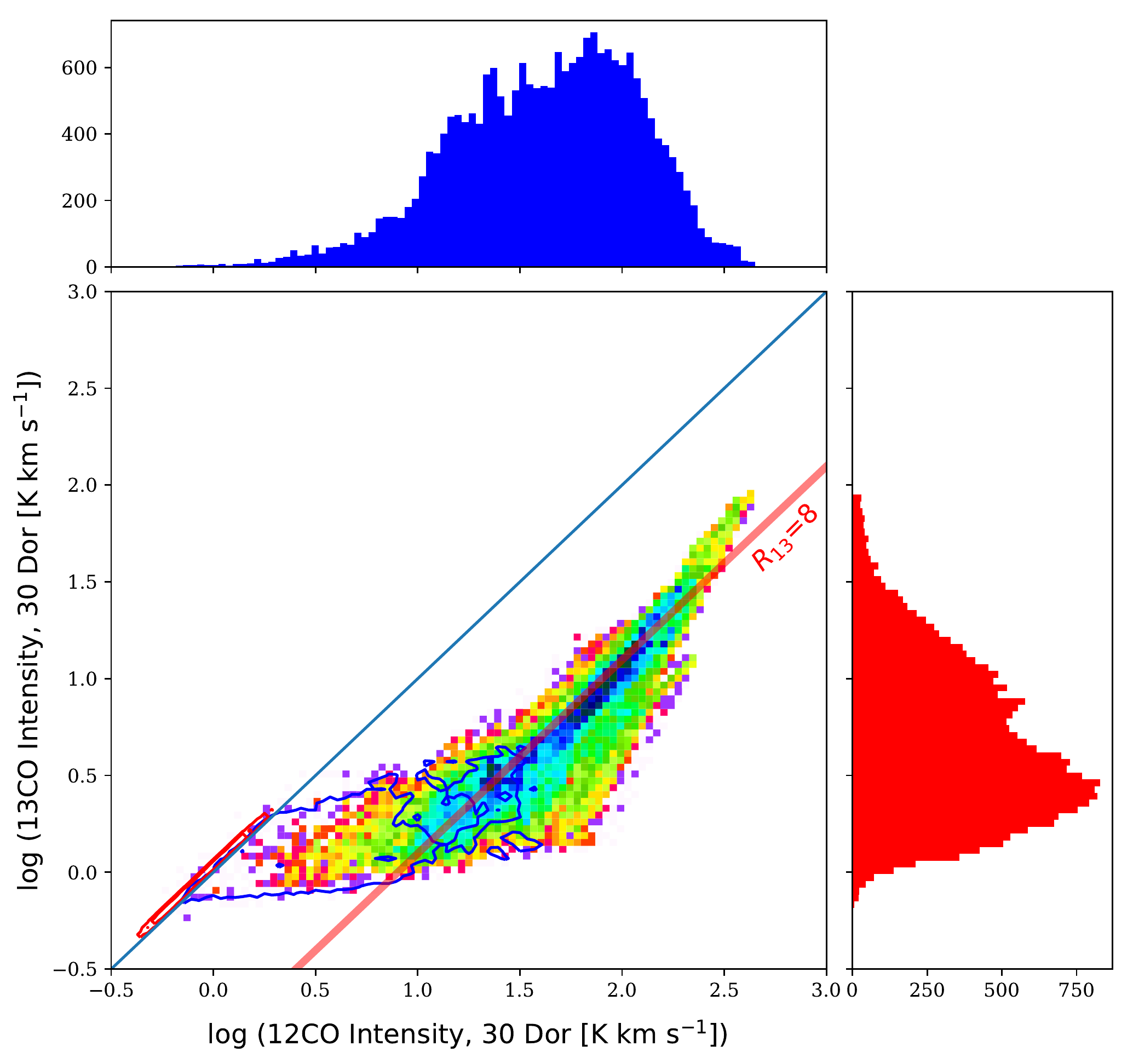}\\
\caption{\small{Correlation between $^{12}$CO and $^{13}$CO integrated intensity for the PCC ({\it left}) and 30Dor cloud ({\it right}).  The comparison is limited to the region enclosed by the dilated $^{12}$CO mask.  Blue contours represent 4$\sigma$ upper limits on $^{13}$CO but with $^{12}$CO detected at a 4$\sigma$ level or greater, whereas red contours are 4$\sigma$ upper limits on both $^{12}$CO and $^{13}$CO.  Blue and red diagonal lines correspond to $I_{12}/I_{13}$ ratios of 1 and 8 respectively. 
}}
\label{fig:i12i13}
\end{figure*}

We derived an LTE-based estimate of clump masses by scaling $N$($^{13}$CO) to $N$(H$_2$) using an abundance ratio of H$_2$/$^{13}$CO of $5 \times 10^6$, to maintain consistency with \citet{Indebetouw:13}.  We compare the LTE-based masses with masses derived from the CO luminosity or virial theorem in Figure~\ref{fig:masscorr}.  The top panels show, for the PCC and 30 Dor respectively, the LTE mass derived by integrating the scaled $N$($^{13}$CO) cube within the boundaries of the \twco-defined structures; this mass is typically $\sim$3 times larger than the CO-based mass.  The bottom panels show the LTE mass integrated within the boundaries of the \ttco-defined structures, compared with the virial mass derived from the \ttco\ sizes and line widths.  The LTE mass overestimates the virial mass by a factor of $\sim$2 in 30 Dor but a factor of 6--8 in the PCC.  For all cases the relation is very close to linear: the LTE masses agree with the other mass estimators aside from a constant multiplicative offset.

One reason for our systematically high LTE masses, besides the underestimate of $T_{\rm ex}$ described above, may be our adopted H$_2$/$^{13}$CO ratio, which is somewhat higher than that adopted in most previous studies \citep[e.g.,][]{Minamidani:11}. The gas-phase [$^{12}$C/$^{13}$C] ratio in the LMC was estimated to be between 30 and 75 by \citet{Johansson:94}, and a value of 36$\pm$11 was derived by \citet{Heikkila:99}.  The conventional value for the H$_2$/CO ratio in Galactic studies is 1--2 $\times 10^4$ \citep{Frerking:82,Blake:87}, but should be somewhat higher in the LMC, with a metallicity of roughly half solar \citep{Stasinska:98}.  Taken together, it is plausible that our H$_2$/$^{13}$CO abundance ratio is overestimated by a factor of 2--7, which would be adequate to explain the discrepancies between the LTE-based masses and the other mass estimators.  Differences in the dynamical state of the PCC and 30 Dor clouds may also systematically affect the virial mass estimates.

Departures from the LTE approximation constitute an additional source of systematic uncertainty,
for instance if photon trapping enhances the population of upper-level \twco\ states above what would be expected from LTE.  This creates a range of densities ($\lesssim 10^3$ cm$^{-3}$, \citealt{Indebetouw:13}) for which \ttco\ is subthermally excited while \twco\ is thermalized.  \cite{Heyer:09} show that the LTE approximation can underestimate the \ttco\ column density by factors of 2 or more in low-density cloud envelopes, with selective photodissociation also reducing the \ttco\ abundance in UV-exposed regions, leading to a further underestimate of the H$_2$ column density.  To diagnose the presence of such diffuse cloud envelopes, we display the pixel-by-pixel correlation of the integrated \twco\ and \ttco\ line intensities in Figure~\ref{fig:i12i13}.  Points at the bottom and left of each plot, below a 4$\sigma$ significance, have been suppressed and replaced by contours indicating upper limits.  The blue diagonal line represents $R_{13} \equiv I(\twco)/I(\ttco) = 1$ and the red diagonal line represents $R_{13} = 8$.  While a range of $R_{13}$ values is observed in both clouds, at our sensitivity there is little evidence for optically thin gas ($R_{13} \gg 10$).
Coupled with the fact that the LTE masses tend to be {\it higher} than other mass estimates, it seems unlikely that diffuse cloud envelopes contribute significantly to the emission we measure.  Rather, the underestimate of $T_{\rm ex}$ by assuming a filled beam and optically thick \twco\ emission, in addition to the somewhat high value of the H$_2$/$^{13}$CO ratio we adopt, is likely biasing the LTE masses high.

\begin{figure*}[tbh]
\begin{center}
\includegraphics[height=3.2in,viewport=7 1 168 282,clip=true]{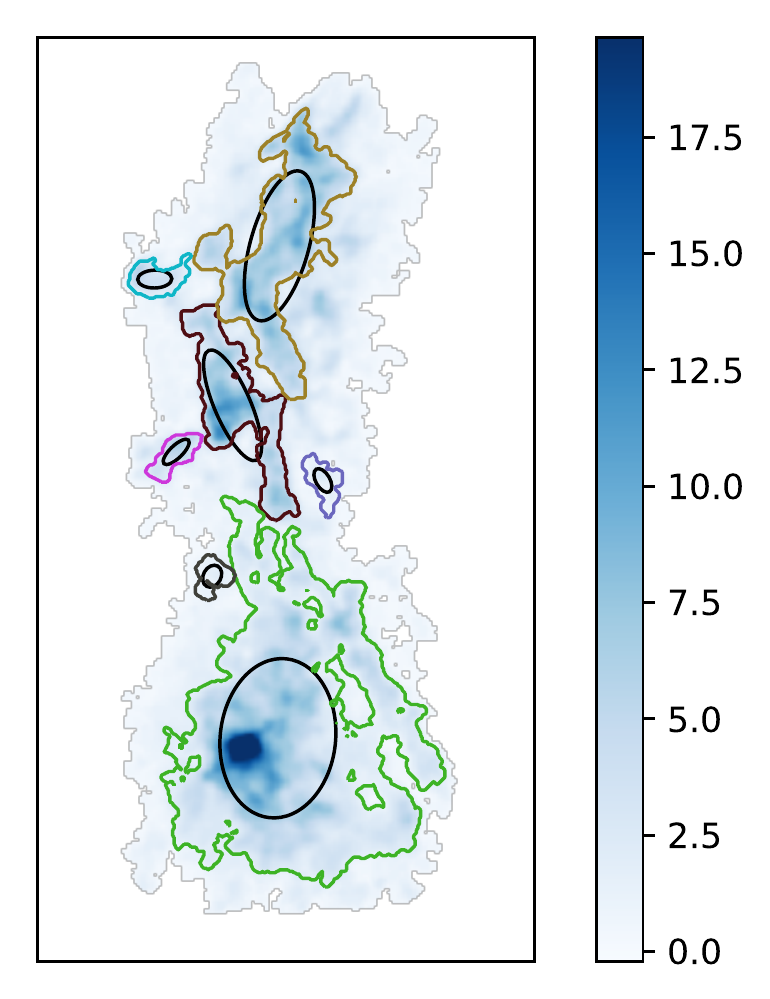}
\includegraphics[height=3.2in]{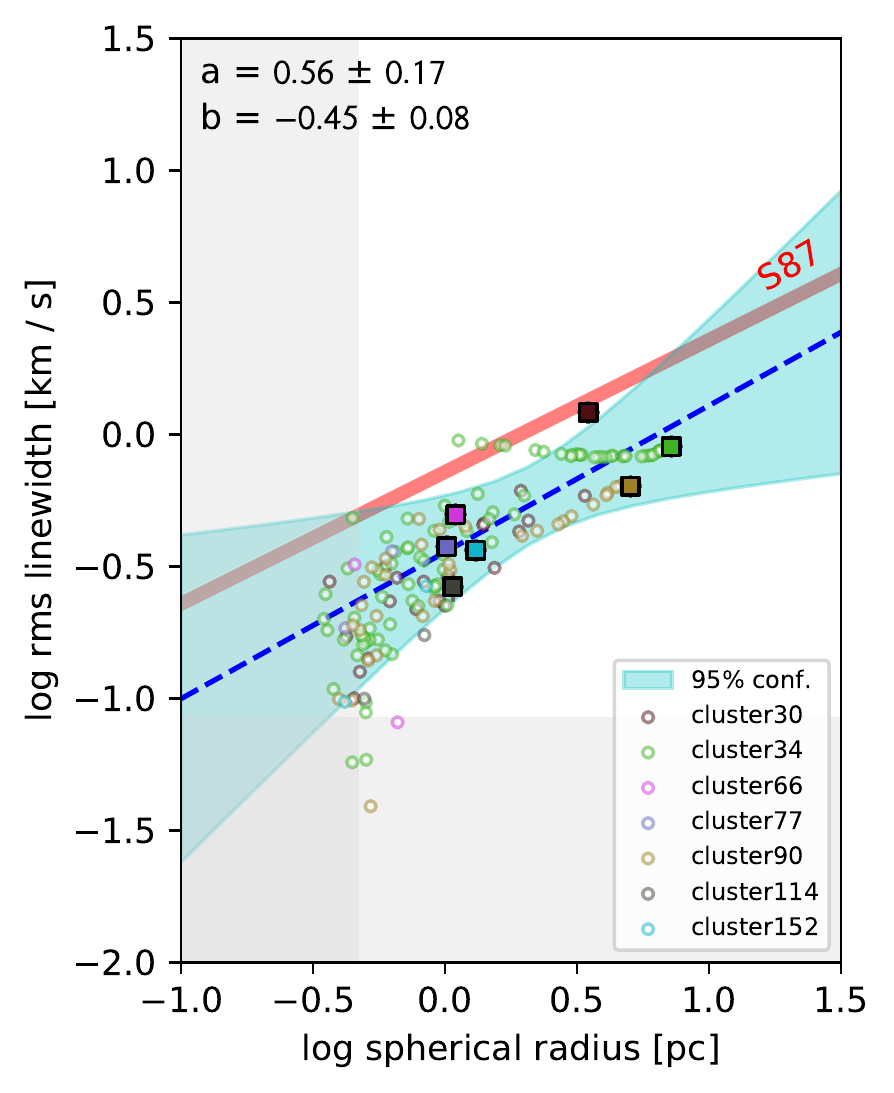}
\includegraphics[height=3.2in]{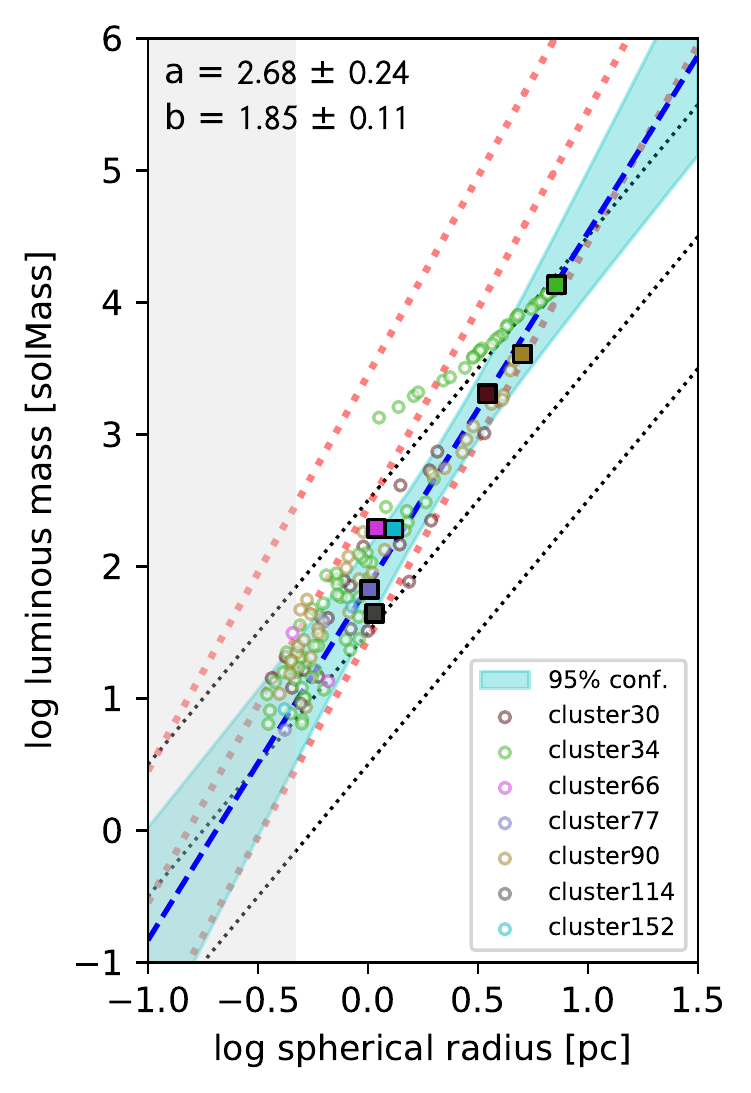}
\end{center}
\caption{\small{{\it Left}: Boundaries (colored contours) and RMS sizes (denoted by black ellipses) of clusters identified in the CO dendrogram of the PCC using SCIMES.  {\it Middle}: Size-linewidth relation for clusters (colored squares, using the same color key as the left panel).  Substructures within each cluster are shown as open circles with the same color as the cluster.  The power-law relation and its uncertainty are determined based on the cluster points only.  The Galactic relation from \citet{Solomon:87} is shown as a red line.  {\it Right}: Size-mass relation for clusters (colored squares).  Dotted lines are the same as in Fig.~\ref{fig:areaflux}.
}}
\label{fig:clustermap}
\end{figure*}

\section{Discussion}\label{sec:disc}

\subsection{The Southern IR Source}\label{sec:core}

The bright, compact CO source in the southern part of the PCC is associated with an infrared (IR) point source seen in the SAGE and HERITAGE surveys of the LMC, with a 24\,$\mu$m flux of 4.87$\pm$0.09 mJy \citep{Seale:14}.
In conjunction with the unusually broad line profile seen in the CO spectra (Fig.~\ref{fig:dendro} and \ref{fig:spectra}), this is indicative of a luminous young stellar object (YSO) or compact cluster that is driving a powerful wind or outflow.
\citet{Seale:14} obtain a far-IR luminosity of 330$\pm$74 $L_\odot$ for this source (designated J81.048316$-$71.913885) based on graybody fitting to HERITAGE photometry.
We further combined the available 2MASS, SAGE, and HERITAGE photometry with SED fitting to the models of \citet{Robitaille:07} to estimate a total luminosity of $\sim$1000 $L_\odot$.
Although the quality of the fit is poor, it implies a mass of $\sim 8\,M_\odot$, an extinction $A_V \sim 100$, and an envelope accretion rate of $\sim 3 \times 10^{-3}$ $M_\odot$ yr$^{-1}$.
Thus, despite the low temperature and IR brightness of the PCC, star formation does appear to be underway: the absence of a bright \HII\ region does not imply a lack of energetic feedback.
Still, the upper left panel of Fig.~\ref{fig:dendro} suggests that the vicinity of the YSO has not been the site active star formation until recently ($\lesssim 10^6$ yr), as structures within a few pc of the IR source do not show elevated line widths.

\subsection{Variations in line width on small scales}\label{sec:scimes}

A key question in star formation theory remains the source and maintenance of molecular cloud turbulence \citep[e.g.,][]{Klessen:10}.  Comparison with simulations has suggested that the observed $R$--$\sigma_v$ relations in Galactic molecular clouds are more consistent with large-scale driving (i.e., on scales comparable or larger than the cloud size) than with small-scale driving from stellar feedback \citep{Ossenkopf:02,Brunt:09}.  Two of our principal results, however, suggest that internal feedback effects may have significant impact on observed cloud properties and scaling relations.

First, we find clear evidence for small-scale energy injection in the PCC, at the location of the southern IR source.  Indeed, we have noted that the ``leaf'' structures appear to span nearly the full range of velocity dispersions observed in structures of all scales (Fig.~\ref{fig:rdv}).  
In terms of total cloud kinetic energy, the leaves still play only a minor role, consistent with the majority of them being quiescent. Approximating the kinetic energy as $M_{\rm lum}\sigma_v^2$, the leaves as a whole contribute only 5.5\% and 5.1\% of the energy in PCC and 30 Dor respectively, compared to 18\% and 32\% of the mass.
Notably, however, the leaf with the highest kinetic energy in the PCC makes similar contributions to the mass (4.3\%) and kinetic energy (3.9\%) of the cloud as a whole.
The most energetic leaf in 30 Dor is less dominant among leaves (contributing 4.1\% of the mass but only 1.7\% of the energy) but still provides a third of the kinetic energy of all the leaves in total.  These results suggest that energy injection on small scales is likely to be highly localized in space and time, increasing the scatter in observed size-linewidth relations.

Second, we have found clear differences in line width for two clouds (PCC and 30 Dor) with very different star formation activity, challenging the notion that star formation activity has little effect on the strength of turbulent motions in a cloud.  Of course, given the many peculiarities of the 30 Dor region (in particular, its extremely high radiation field, e.g.\ \citealt{Lopez:11}), the properties of this cloud may prove to be extreme, and better context will be available with upcoming observations of many more clouds.

To better understand why previous analyses have often failed to uncover large variations in line width on small scales, we can compare our results to a similar analysis performed using SCIMES, an extension to {\it astrodendro} that segments the emission into closely related structures via cluster analysis \citep{Colombo:15}.  A SCIMES decomposition results in distinct, non-overlapping structures that resemble the clumps identified by CLUMPFIND \citep{Williams:94} or CPROPS \citep{Rosolowsky:06}.  Since SCIMES works directly on the dendrogram output, we can relate the ``clusters'' it identifies with the full set of structures in the dendrogram.  We find, for the PCC \twco\ cube, that SCIMES breaks the emission into just 7 clusters, the largest of which includes the southern IR source (Figure~\ref{fig:clustermap}).  Because small, high-linewidth structures like the southern CO-bright ``leaf'' tend to lie well above the noise floor, they are preferentially eliminated by incorporation into larger structures.
The resulting clusters tend to lie closer to a power-law relation than when the full set of dendrogram structures are included, although the slope is formally more uncertain because of the small number of points.

The possibility that clouds derive a significant fraction of their turbulent energy from internal feedback has been noted by \citet{Goldbaum:11}, who studied the stability and lifetimes of clouds experiencing both mass accretion and star formation.  Yet it remains unclear whether stellar feedback can provide turbulent support against gravity on the $\gtrsim$10 pc scales of GMCs (supernovae could do so easily, but would likely disrupt the cloud).  The presence of bulk velocity gradients in Galactic and extragalactic GMCs \citep{Koda:06,Imara:11} suggests that both feedback and cloud-scale bulk motions can provide support to varying extents.  Recognizing a role for feedback may provide a path for understanding why star formation appears to be enhanced in GMCs that have experienced previous star formation events \citep{Ochsendorf:16}.  With a diverse set of well-resolved GMCs all at a known distance, the LMC will be a key laboratory for further exploration of these issues.

\subsection{High Densities Closer to Virial Equilibrium}\label{sec:key}

The comparisons we have made between the PCC and 30 Dor clouds present an interesting conundrum.  While the basic scaling relations between size and line width (Fig.~\ref{fig:rdv}) and size and mass (Fig.~\ref{fig:areaflux}) have slopes consistent with previous studies, there is a clear difference in normalization between the two clouds.
At the same time, the comparisons between virial and luminous masses and surface densities in Fig.~\ref{fig:mlumvir} and \ref{fig:bnd} show that the highest density structures in both clouds remain close to virial equilibrium.  While 30 Dor displays much higher line widths and surface densities than PCC, the normalization of the size-linewidth relation ($\sigma_v/R^{0.5} \propto \sqrt{\Sigma_{\rm vir}}$) scales with surface density, roughly as expected from virial equilibrium.  This scaling appears to explain much of the difference in clump properties between the PCC and 30 Dor, and is even more striking when comparing structures within the PCC, where higher surface density clearly implies higher line width at a given size (Fig.~\ref{fig:rdvcolor}).

It is worth bearing in mind, however, that many of the smaller structures in 30 Dor appear well out of equilibrium, with virial masses exceeding luminous masses by an order of magnitude or more.  
Although, as noted in \S\ref{sec:dendro}, our ``bijection'' approach to measuring cloud properties likely underestimates the line widths of smaller structures, correcting for this bias would only increase the discrepancy.
Given that $\alpha_{\rm vir} = M_{\rm vir}/M_{\rm lum} = 2{\cal T}/|{\cal W}|$, the observed offset implies a strong excess of kinetic energy in these structures, or the presence of a confining external pressure.  Interestingly, the brightest PCC clump associated with the southern IR source in that cloud has $\alpha_{\rm vir} \approx 1$.  This suggests that even strong internal feedback need not lead to large departures from virial equilibrium, for example if internal feedback ``stirs'' the dense gas but is unable to completely unbind it.  The highly supervirial line widths seen in parts of the 30 Dor cloud may instead result from {\it external} kinetic energy injection (e.g., from the nearby R136 cluster) that is not limited by local escape speeds, or by a significant stellar contribution to the gravitational potential that increases the virial line width above what would be expected from the gas mass alone (cf.\ discussion in \citealt{Kruijssen:13}).
Alternatively, the majority of the apparently unbound structures could be transient density enhancements and thus not in dynamical equilibrium at all \citep[e.g.,][]{Dib:07}.

\section{Conclusions}\label{sec:conc}

We have presented Mopra and ALMA observations of \twco(1--0), \twco(2--1), and \ttco(2--1) emission from a quiescent cold cloud in the outskirts of the LMC.\@  We have confirmed that the cloud has among the coldest dust temperatures for CO clouds in the galaxy and possesses a relatively high \twco(2--1)/\ttco(2--1) line ratio but low \twco(2--1)/\twco(1--0) line ratio, consistent with lower density and temperature than is typical for more centrally located clouds.  Comparing the PCC cloud with the well-studied 30 Dor cloud using the same analysis techniques at the same physical resolution, we find the following:

\begin{enumerate}

\item Line widths in 30 Dor are larger than line widths in the PCC by a factor of $\sim$5 at a given spatial scale.  As a result, the best-fit $R$-$\sigma_v$ relations for each cloud straddle the average relation found in Galactic clouds by \citet{Solomon:87}.

\item There is a large range in line widths measured for the ``leaves'' that occupy the smallest scales in the dendrogram hierarchy.  Especially in the PCC, but to some extent also in 30 Dor, the velocity dispersion observed on larger scales is heavily influenced by the highest dispersion leaves, resulting in a relatively flat upper envelope to the size-linewidth relation.

\item The CO surface brightnesses and inferred surface densities in 30 Dor exceed those in the PCC by an order of magnitude, for structures of the same size observed at a resolution of 0.6 pc.  

\item The highest density structures in both clouds are close to virial equilibrium, with higher surface densities in 30 Dor compensated by higher line widths.  However, many of the smaller structures, particularly in 30 Dor, appear far out of equilibrium, indicating excess kinetic energy or a confining (but spatially varying) external pressure.

\item Mass estimates based on \twco\ brightness, simple virial equilibrium, and \ttco\ emission in LTE show strong correlations, yet systematically differ by up to a factor of $\sim$6.  These estimates each rely on assumptions that may not be valid for individual clouds, and should be treated with caution.

\end{enumerate}

The clear and dramatic differences between PCC and 30 Dor illustrate the dependence of molecular cloud properties on galaxy environment, a topic which has been the focus of significant attention in recent years \citep[e.g.,][]{Hughes:13b,Fujimoto:14,Freeman:17}.  The 30 Dor-10 cloud is located in an extreme star-forming environment where young stars have dispersed much of their natal gas; it is also in a more central location in the galaxy than the PCC.  High surface densities and velocity dispersions, such as observed in 30 Dor, translate into high internal cloud pressures ($P_{\rm int} \sim \Sigma \sigma^2/R$), which have been found to correlate with the external pressure required for hydrostatic equilibrium of the galactic disk \citep{Hughes:13b}.  While it is to be expected that the properties of clouds reflect to some degree the properties of the environment from which they form, the implications of GMC properties for star formation are only beginning to be explored \citep[e.g.,][]{Kruijssen:14,Federrath:16,Ochsendorf:17}.  In future work, we will revisit these topics with a more comprehensive survey of molecular clouds in the LMC, which probe a more continuous range of environmental conditions.

\acknowledgments
The Mopra radio telescope is part of the Australia Telescope National Facility which is funded by the Australian Government for operation as a National Facility managed by CSIRO.
This paper makes use of the following ALMA data: ADS/JAO.ALMA 2011.0.00471.S, 2013.1.00832.S. ALMA is a partnership of ESO (representing its member states), NSF (USA) and NINS (Japan), together with NRC (Canada), NSC and ASIAA (Taiwan), and KASI (Republic of Korea), in cooperation with the Republic of Chile. The Joint ALMA Observatory is operated by ESO, AUI/NRAO and NAOJ.
The National Radio Astronomy Observatory is a facility of the National Science Foundation operated under cooperative agreement by Associated Universities, Inc.
AH acknowledges support from the Centre National d'Etudes Spatiales (CNES).
ER is supported by a Discovery Grant from NSERC of Canada.
KT was supported by NAOJ ALMA Scientific Research Grant Number 2016-03B.
The work of M.S. was supported by an appointment to the NASA Postdoctoral Program at the Goddard Space Flight Center, administered by Universities Space Research Association under contract with NASA.
Part of this research was conducted at the Jet Propulsion Laboratory, California Institute
of Technology, under contract with the National Aeronautics and Space Administration.
This research made use of {\it astrodendro}, a Python package to compute dendrograms of astronomical data, SCIMES, a Python package to find relevant structures in dendrograms of molecular gas emission using the spectral clustering approach, and Astropy, a community-developed core Python package for astronomy.

\facility{Mopra, ALMA, Planck, Herschel, Spitzer}

\software{CASA \citep{McMullin:07}, DUSTEM \citep{Compiegne:11}, {\it astrodendro} (\url{http://www.dendrograms.org}), Kapteyn \citep{KapteynPackage}, Scipy \citep{Jones:01}, SCIMES \citep{Colombo:15}, Astropy \citep{Astropy:13}.}

\bibliographystyle{aasjournal}
\bibliography{merged}

\end{document}